\documentclass[useAMS,usenatbib,usegraphicx]{mn2e}

\usepackage{amsmath}
\usepackage{amssymb}

\def\deg{\hbox{$^\circ$~\/}}
\def\degn{\hbox{$^\circ$\/}}
\def\arcminn{\hbox{$^\prime$\/}}

\def\Msun{\hbox{$\rm M_{\odot}~$}}

\def\Msunn{\hbox{$\rm M_{\odot}$}}
\def\Zsunn{\hbox{$\rm Z_{\odot}$}}

\title[\textit{XMM-Newton} observations of the Galactic Centre Region
  - II]{{\it XMM-Newton} observations of the Galactic Centre Region -
  II: The soft thermal emission}
\author[V. Heard and
  R. S. Warwick]{V. Heard$^{1}$
\thanks{E-mail: vh41@le.ac.uk} and
  R. S. Warwick$^{1}$\\ $^{1}$Department of Physics and Astronomy,
  University of Leicester, University Road, Leicester, UK}

\begin{document}

\date{Accepted . Received ; in original form}

\pagerange{\pageref{firstpage}--\pageref{lastpage}} \pubyear{2013}

\maketitle

\label{firstpage}

\begin{abstract}

\noindent  We have extended our earlier study (\citealt{heard13},
Paper I) of the X-ray emission emanating from the central 100 pc $\times$ 100 pc
region of our Galaxy to an investigation of several features prominent in the
soft X-ray (2--4.5 keV) band. We  focus on three specific
structures:  a putative bipolar outflow from the vicinity of Sgr  A*;
a high surface brightness region located roughly 12 arcmin (25 pc)
to the north-east of Sgr A*; and a lower surface-brightness extended
loop feature seen to the south of Sgr A*.
We  show,  unequivocally, that all three structures  are thermal  in
nature and have similar temperatures ($kT \approx 1$ keV). The
inferred X-ray luminosities lie in the range 
$(2 - 10) \times 10^{34} \rm~erg~s^{-1}$.
In the case of the bipolar feature we suggest  that the hot plasma is produced
by  the shock-heating  of  the  winds from  massive  stars within  the
Central Cluster, possibly collimated  by the Circumnuclear Disc.  
Alternatively  the  outflow   may  be driven by outbursts on  Sgr A*,
which follow tidal disruption events occurring at a rate of roughly 1
every 4000 yr.  The   north-east enhancement is centred on
a candidate pulsar wind nebula which has a relatively hard
non-thermal X-ray spectrum. We  suggest that 
the coincident soft-thermal emission traces the core of a new
thermal-composite supernova remnant, designated as
SNR G0.13-0.12. There is no clear evidence for an associated radio shell
but such a feature may be masked by the bright emission of the nearby
Radio Arc and other filamentary structures. SNR G0.13-0.12 is very likely
interacting with the nearby molecular cloud, G0.11-0.11, and linked to
the \textit{Fermi} source, 2FGL J1746.4-2851c. Finally
we explore a previous suggestion that the elliptically-shaped X-ray loop
to the south of Sgr A*, of maximum extent $\sim$ 45 pc, represents the
shell of a superbubble located in the GC region. Although plausible,
the interpretation of this feature in terms a coherent physical
structure awaits confirmation.

\end{abstract}


\section{Introduction}
\label{sec:intro}

The central 100-pc of our Galaxy is a fascinating region which abounds
with unusual  astrophysical features and exotic  phenomena. The region
is  characterised   by  a  high   concentration  of  both   stars  and
interstellar  material with  an the  enclosed mass  of $\sim  3 \times
10^{8}$ \Msun  (\citealt{launhardt02}).  The non-stellar  component is
distributed in  dense molecular clouds  within a structure  known as
the Central Molecular Zone  (\citealt{morris96}; \citealt{molinari11}).  
The presence within the
region of three of the most  massive young star clusters in the Galaxy
(\citealt{figer99}; \citeyear{figer02}; \citealt{paumard06}) is just one of
many  signatures of  recent star  formation.  An  enhanced rate  of
massive star  formation gives rise  to supernova (SN) explosions every few
thousand  years and  hence  to a  ready supply  of  heat and  accelerated
particles (\citealt{crocker11}).
These  in turn  fuel  a  range  of energetic  thermal  and
non-thermal  emission  processes  ({\it  e.g.,} \citealt{koyama89}; 
\citealt{yamauchi90}; \citealt{belanger04}; \citealt{muno04};  
\citealt{crocker11}; \citealt{yusef13}).
A supermassive  black  hole
(SMBH)   of  mass   $(4.5~\pm 0.4)   \times  10^{6}~\Msunn$   
(\citealt{schodel02}; \citealt{ghez08}, \citealt{gillessen09}) 
resides at  the  dynamical centre  of the  Galaxy;
although  currently in  a  quiescent accretion  state,  the SMBH  may,
during past  outbursts, have further energised  the surrounding region
(\citealt{koyama96};      \citealt{murakami11};     \citealt{ponti10};
\citealt{nobukawa11}; \citealt{capelli12}; \citealt{gandoryu12}).
Accretion power also  drives the luminous X-ray
binaries and transient  sources which are very prominent  in X-ray and
$\gamma$-ray observations of the GC (\textit{e.g.}, \citealt{sidoli99a}; 
\citealt{muno06}; \citealt{degenaar12}).

In an earlier paper (\citealt{heard13}, hereafter Paper I), we
used  \textit{XMM-Newton}  observations  in  an investigation  of  the
spatial  distribution  and spectral  characteristics  of the  extended
X-ray  emission  associated with  the  central  100-pc region.   More
specifically,  we argued  that the  very hot  thermal emission
which is particularly bright in  the GC and which can be characterised
spectrally  as a  thermal bremsstrahlung  component ($kT  \approx 7.5$
keV), plus  attendant He-like and  H-like iron K-shell lines,  is best
explained  in terms  of the  integrated emission  of  unresolved point
sources.  It seems likely that  this population of X-ray sources (with
the X-ray  luminosity, $L_{\rm{X}}$, of  individual sources  typically
in the   range   $10^{30}-10^{33}   \rm~erg~s^{-1}$)  is   dominated   by
cataclysmic  variables (CVs) (\citealt{muno06}; \citealt{revnivtsev11};
\citealt{yuasa12}). In  Paper  I we  found  that a  smooth,
centrally-concentrated spatial   model  for   these
unresolved sources  accounted for the  bulk of the He-like  and H-like
iron line emission at 6.7 keV  and 6.9 keV respectively, 
without recourse  to a  significant contribution  from a  very-hot
{\it truly diffuse} X-ray plasma.  
The underlying iron fluorescent line emission at  6.4 keV was also
broadly accounted for by the unresolved-source model,  once the
bright fluorescing knots associated with specific dense molecular
clouds were excluded.

However,  this  is not  the  full story  since  the  GC diffuse  X-ray
emission   also  exhibits  prominent  spectral  lines  due  to
K-shell transitions in highly-stripped  ions of elements such as Si,
S and  Ar, in ratios which can be readily explained in  terms of thermal
plasma emission at temperatures in the range 0.6--1.5 keV
({\it e.g.,} \citealt{koyama89}, \citeyear{koyama96}; \citealt{kaneda97}; 
\citealt{tanaka00}; \citealt{muno04}; \citealt{nobukawa10}), 
{\it i.e.,} a  temperature significantly lower than that characterising
the  spectrum of the unresolved sources.  The spatial  distribution of this
cooler plasma is far from  uniform,  consistent with  the  hypothesis
that  it is  largely generated through  the interaction of SN shock
fronts with the ambient interstellar matter.
 
In this  paper (Paper II), we  extend our study of  the X-ray emission
emanating from the central 100-pc  of the Galaxy to a consideration of
several regions  which are very  bright in the  soft X-ray band,  as a
consequence  of  their  soft-thermal  emission.  We  concentrate  on  three
features:  (i) the ``bipolar'' emission complex which extends 6 arcmin (14 pc) 
each side of Sgr A* along a perpendicular to the Galactic Plane; 
(ii) the  bright emission 
located between  $8-15$ arcmin ($18-35$ pc) to the
north-east\footnote{In this paper directions are referenced
to the celestial coordinate system unless otherwise specified.} 
of  Sgr A*, centred on the pulsar wind nebula (PWN)
G0.13-0.11; and (iii) a putative superbubble of dimension 
$25\arcminn \times 16\arcmin$ (45 pc $\times$ 36 pc)
located to the south of Sgr A*.
In this  study  we attempt  to characterise  the
properties and  spatial distribution  of the soft  X-ray emitting plasma
in  these three  regions  as a step towards charting  the activity  and
processes which have helped shape the extreme GC environment.

In the next section (\S2) we briefly summarise the \textit{XMM-Newton}
observations used  in our study  and the data reduction  techniques we
have  employed. In  \S3, we  present an  image of  the central  100-pc
region in  the 2--4.5 keV band  and, through the  use of
latitudinal  cuts across  images constructed  in  different wavebands,
take an overview  of the properties of the  soft X-ray emitting plasma
present  in the region.  The next  three sections  (\S4, \S5  and \S6)
report  a more  detailed  analysis  of the  three  X-ray bright  regions
identified  above.    Finally  in  \S7  we summarise our results
and conclusions. Throughout this  work, the distance to the  GC is
assumed to be 8 kpc \citep{gillessen09}.

\section{Observations and data reduction}
\label{sec:obs}

\subsection{Image construction and analysis}
\label{sec:imagecon}

Full details of the \textit{XMM-Newton}  observations used in this analysis
and a description  of the techniques employed in  the reduction of the
data    from   the    three    EPIC   cameras    (\citealt{struder01};
\citealt{turner01}) can be found in Paper I.  In brief, a set of 50 pn
and  56 MOS-1/MOS-2 observations,  giving a  total useful  exposure of
$\sim0.5$  Ms and  $\sim0.9$  Ms respectively,  were  used in  the
construction  of a  set of  mosaiced images  centred on  (RA,  Dec) $=$
(266.41657$\degn$, -29.0787$\degn$). The image format comprised $648
\times 648$ pixels (4 arcsec pixels), corresponding to a $43.2\arcminn
\times 43.2\arcmin$ field, which at  the distance of the GC equates to
a projected extent of 100 pc $\times$ 100 pc.

For this work, the mosaiced  images were constructed in five ``broad''
bands covering  a nominal  total bandpass from  $1-10$ keV  (see Table
\ref{tab:bands}). In  addition,  images  were   made  in  a  number  of
``narrow''  bands  matching  prominent  spectral   lines.   Table
\ref{tab:bands}  lists  the  narrow-band  energy ranges  used  in  the
present  study.  These correspond  to  the  K-shell  lines of  He-like
silicon, sulphur and  argon, the neutral  (or near-neutral)
iron fluorescence line  at 6.4 keV (Fe \textsc{i}  K$\alpha$), and the
K-shell lines  at 6.7  keV and 6.9  keV from He-like  (Fe \textsc{xxv}
K$\alpha$)  and  hydrogenic  (Fe  \textsc{xxvi}  Ly$\alpha$) ions of
iron. Narrow-band  images sampling  a spectral range  near to  that of
each emission line were  also produced (see Table \ref{tab:bands}) and
used to subtract the continuum underlying each line (see below).

\begin{table}
\begin{center}
\caption{The energy ranges (in keV) of the ``broad'' bands covering
the 1--10 keV bandpass and the ``narrow'' bands encompassing  specific
spectral lines.}  \renewcommand{\arraystretch}{0.75} \small
\begin{tabular}{c l}
\hline    
\multicolumn{2}{c}{Broad bands}\\   
\hline
very-soft&$1.0-2.0$\\
soft&$2.0-4.5^{\dagger}$\\
hard&$4.5-6.0$\\  
iron&$6.0-7.2$\\  
very-hard&$7.2-10.0^{\ddagger}$\\  
\hline
\multicolumn{2}{c}{Narrow bands}\\  
\hline
Si \textsc{xiii} K$\alpha$  &$1.740-1.920$\\
Cont. Si  &$1.920-1.965$ \\
... &$2.070-2.115$\\
S \textsc{xv} K$\alpha$  &$2.340-2.550$\\
Cont. S  &$2.070-2.115$\\
... &$2.250-2.310$\\
... &$2.700-2.805$\\
Ar \textsc{xvii} K$\alpha$  &$3.015-3.225$\\
Cont. Ar  &$2.910-3.015$\\
... &$3.225-3.330$\\
Fe \textsc{i}    K$\alpha$  & $6.270-6.510$\\  
Fe \textsc{xxv}  K$\alpha$ & $6.525-6.825$\\  
Fe \textsc{xxvi} Ly$\alpha$ & $6.840-7.110$\\  
Cont. Fe low&  $5.700-6.150$\\
Cont. Fe high& $7.200-7.650$\\
\hline
\end{tabular}
\label{tab:bands}
\end{center}
$^{\dagger}$ In the equivalent table in Paper I, the 2--4.5 keV bandpass was
referred to as the medium (energy) band.\\
$^{\ddagger}$The $7.8-8.3$ keV spectral region encompassing the Cu K$\alpha$
instrumental line was excluded in the case of the pn data, whereas for
the MOS data the very-hard band was truncated at 9.0 keV. 
\end{table}

Paper I  describes how we  constructed mosaiced images  separately for
the  pn and  MOS  (MOS 1  + MOS  2)  cameras; however,  for the  image
analysis  in  this  paper,  all  the data  have  been  co-added.   The
procedure for combining the pn and MOS channels was as follows.  First
the  ratio  of the  MOS  to  pn count  rate  was  calculated for  each
broad band.  This  was  the  field-average  value  determined  from  a
comparison of  the separate pn  and MOS images.  For  each broad band,
the pn,  MOS 1 and  MOS 2 counts  from each observation  were co-added
into a combined-counts image  formatted as described above.  Similarly
a combined-exposure map was constructed,  where both the MOS 1 and MOS
2 contributions  were scaled  by the (single  camera) MOS to  pn count
rate ratio appropriate to the  band.  The final flat-fielding step for
the broad  bands then involved dividing each  combined-counts image by
the  corresponding   combined-exposure  map.   In  the   case  of  the
narrow-band image centred on the He-like Si line, we used  the $1.0-2.0$ keV
exposure map in this flat-fielding  step.  Similarly,  for  the narrow-band
images centred on the He-like S and  Ar lines,  the $2.0-4.5$  keV
exposure map  was  utilised. Finally for the narrow-band images centred
on the three iron lines, we used the $6.0-7.2$ keV exposure map.
Hereafter we refer to the narrow-band images simply as the Si, S,
Ar, Fe64, Fe67 and Fe69 images. 

In  order  to trace  the  distribution of  the  line  emission, it  is
necessary  to make  a correction  for  the
in-band continuum emission underlying  the emission lines. For the Si,
S and  Ar images, a straight subtraction  of the appropriate nearby
continuum narrow-band  image (after applying  a suitable scaling  for the
sampled bandwidth) proved an  adequate approximation.  See Paper I for
a  discussion of  the continuum-subtraction  procedure adopted  in the
case of  the Fe64, Fe67 and Fe69 images and for  the approach used to
correct for the spillover of each iron line into the adjacent narrow bands.

\begin{figure*}
\centering
\begin{tabular}{c}
\includegraphics[width=140mm,angle=0]{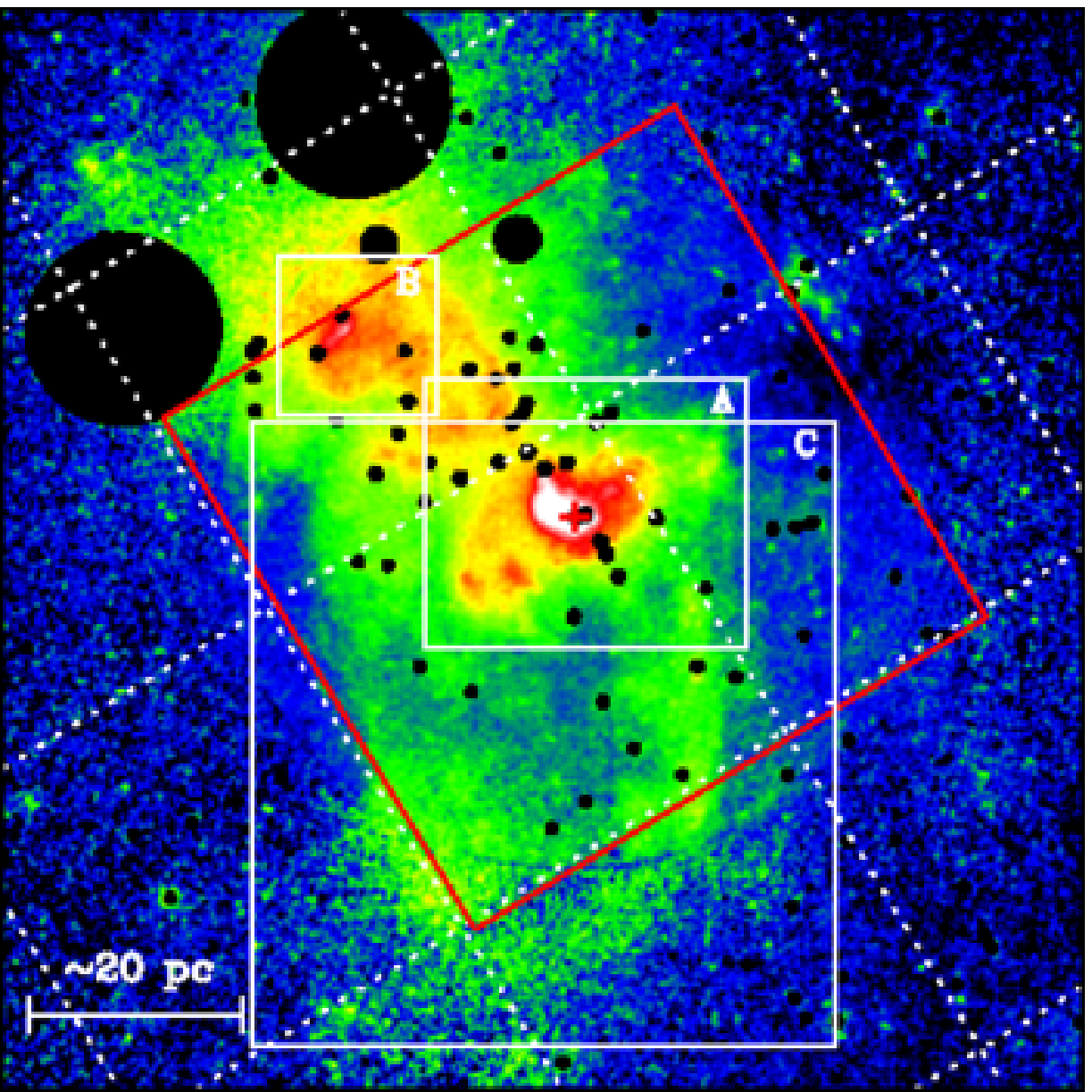}\\
\includegraphics[width=140mm,angle=0]{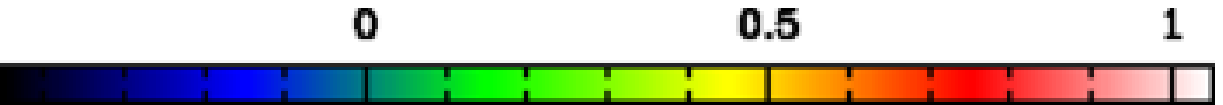}\\
\end{tabular}
\caption{Combined  pn/MOS mosaiced  image  in the  $2.0-4.5$ keV  band
covering the central  43.2\arcminn $\times$ 43.2\arcmin~field of the GC
(100 pc $\times$ 100 pc at the  GC distance). The mosaic is constructed on
a  celestial coordinate  grid with  east  to the  left. The  intensity
scaling  is logarithmic; the  colour bar  shown at  the bottom  of the
figure is calibrated in units of log$_{10}$ (count/20ks/pixel). 
The  image  has  been  spatially smoothed  using  a  circular Gaussian
function of width $\sigma$ = 1 pixel (1 pixel = 4 arcsec). A
source mask has  been applied so as to exclude the  bulk of the signal
from  luminous  discrete  sources  --  see  text. The white dotted lines
represent a Galactic coordinate grid with a 15 arcmin spacing.
The red box centred on  Sgr A* (the position of which is marked by
the red cross)  indicates the region over which  the latitudinal
cuts were calculated  (see \S\ref{sec:cuts}). The white boxes indicate the regions within which individual soft thermal features are contained (see \S\ref{sec:bipolar}--\S\ref{sec:sb}).}
\label{fig:245}
\end{figure*}

By way of illustration, Fig. \ref{fig:245}  shows the combined  pn/MOS
mosaiced image  for the
soft X-ray ($2.0-4.5$ keV) band  after applying both a source mask and
light spatial  smoothing. The  source mask was  designed to  exclude a
significant fraction  ($>$ 80 per cent) of  the signal from  discrete sources
with X-ray  luminosity $> 10^{33}  \rm~erg~s^{-1}$ (2--10 keV).  A few
modifications were made  to the mask employed in  Paper I, namely: (i)
the current  mask excludes the  peak emission of the central
star cluster, which  has Sgr A*  at its  core; however 
it does  not remove  the extended
emission associated with  the Sgr A East SNR;  (ii) extended continuum
emission directly associated with the PWN G0.13-0.11 is not excised; (iii)
faint emission  associated with the  Quintuplet cluster is  now masked
out.  The impact  of the  masking is  most evident  in the north-east
corner of the image, where large excision circles mark the locations of
two very luminous sources, the X-ray transient SAX J1747.0-2853 
(\citealt{wijnands02}; \citealt{degenaar12}) 
and the persistent source 1E 1743.1-2843
(\citealt{porquet03}; \citealt{delsanto06} ). In this same
region the two
mid-size excision circles correspond to the Quintuplet and Arches star
clusters (\citealt{yusef02a}; \citealt{law04}).

\subsection{Spectral extraction and modelling}
\label{sec:spectralx}

For the current work, the X-ray  spectral analysis was largely based on a
single pn  observation (0202670801) for which the  exposure time after
filtering was $\sim$ 62 ks. The one exception to this was the
investigation of G359.79-0.26 (see \S\ref{sec:sb}), which was outside of the
field of view. In this case, we used an alternative observation (0112971001)
for which the exposure time after filtering was $\sim$ 7 ks. 
The \textit{XMM-Newton} \textsc{sas} \textsc{v}11.0 (hereafter, \textsc{sas})
was used to extract the  data and generate all the instrument response
files required for the spectral analysis.  

The wide extent and complex
morphology of  the diffuse soft X-ray  emission in the  GC field makes
the  selection of  a suitable  background region  for use  in spectral
analysis extremely  problematic. We  have therefore followed  the same
approach to background  subtraction as described in Paper  I, and made
use of EPIC filter wheel closed (FWC) data. FWC spectra were extracted
from geometrical  areas on the  pn detector matching those  utilised in
the actual observations  and then scaled by the  ratio of the exposure
times  of the  observation  and FWC  datasets.  The use  of these  FWC
spectra  as  background datasets  proved  fully  satisfactory for  our
application, namely the  study of GC regions of  relatively high, soft
X-ray surface  brightness. One consequence of using FWC data as the
background estimator, is that the applied spectral model should,
in principle, include emission associated with both the
Galactic foreground and the cosmic X-ray background, as well
as with the GC. However, in practice, the contribution of
the cosmic X-ray background was negligible, whereas constraints
of any foreground contribution were provided
by the measurements of the net line-of-sight column density, $N_{\rm{H}}$.

In Paper I, we discussed how the GC X-ray spectra could be modelled in
terms of just three major emission components.  The first is the emission
associated with the unresolved hard X-ray emitting point sources
(with $L_X < 10^{33}$ erg s$^{-1}$, 2--10 keV). Here we follow the same approach as 
in Paper I and represent this component as a bremsstrahlung continuum
plus four Gaussian lines representing the iron fluorescence components,
Fe \textsc{i} K$\alpha$ and K$\beta$, and the K-shell lines of Fe \textsc{xxv} and
Fe \textsc{xxvi}. The temperature of the bremsstrahlung was fixed at
7.5 keV and the absorbing  column density set at $N_{\rm{H}} = 12 \times  10^{22}$ cm$^{-2}$. 
The line equivalents widths were set to 235, 735 and 290 eV
for the neutral Fe K$\alpha$ and the He-like and H-like Fe lines
respectively (these are the average values for the Lat +ve and Long -ve 
region in Paper I, table 5). The normalisation of the neutral Fe K$\beta$ 
line was fixed at 11 per cent of that of the K$\alpha$ line
\citep{koyama09}. The line energies and intrinsic widths were fixed
at the values employed in Paper I.  In the spectral fitting, the only
free parameter was the normalisation of the bremsstrahlung continuum.

The second emission component relates to the reprocessing of incident X-ray photons
in dense molecular clouds in the GC.  The most obvious signature is the
Fe K$\alpha$  fluorescence line at 6.4 keV (\citealt{koyama07b},
\citeyear{koyama09};
\citealt{yusef07}; \citealt{ponti10}; \citealt{capelli12}), with the Thomson 
scattering of the incident X-ray flux
giving rise to an associated hard continuum.  Here we model the
latter as a power-law continuum with a photon index fixed at a value
of $1.8$. The iron fluorescence was represented as a combination of
Fe K$\alpha$ and K$\beta$ emission (from neutral or near-neutral
matter). For this component, the equivalent width of the 6.4 keV line
was fixed at 1600 eV (an average value for the Long +ve region studied
in Paper I). As before, the normalisation of the K$\beta$  line  was
set  equal to 11  per  cent  of that of the  K$\alpha$  value.  
The absorbing  column applied to the power-law continuum 
was fixed at 12 $\times$ 10$^{22}$ cm$^{-2}$.  Again the only free parameter
in the spectral fitting was the normalisation of the continuum.

The third component represents the soft X-ray emission emanating from the diffuse
thermal plasma which pervades the GC region. In the present work, the starting point
for the spectral analysis was to assume conditions of  thermal and ionization
equilibrium and that the soft-thermal emission could be adequately represented by
a single-temperature \texttt{vapec} component in \textsc{xspec} \citep{arnaud96}.
In the spectral fitting the plasma temperature, $kT$,  and normalisation
remained free parameters as did the absorbing column density, $N_{\rm{H}}$.
The relative abundances of Si, S and Ar were also allowed to vary\footnote{
The abundances of all the other metals
were fixed at the solar value, since useful constraints
on these parameters could not be derived from the available data.}.

In  the spectral analysis,
raw spectral  channels were grouped so as 
to give a minimum of 20  counts per  bin.  Uncertainties on the 
derived spectral parameters are quoted  at the ninety per cent
confidence range unless otherwise stated. All spectral modelling was carried
out utilising the X-ray spectral-fitting package 
(\textsc{xspec} \textsc{v}12.7.0, \citealt{arnaud96}). Throughout this paper we assume the solar abundance values of \citet{anders89}.
Further details of the fitting of the spectra extracted from specific
regions are given in later sections.

\section{Soft-thermal emission in the GC}
\label{sec:softem}

\subsection{Overview}
\label{sec:overview}

\begin{figure*}
\centering
\begin{tabular}{cc}
\includegraphics[width=55mm,angle=270]{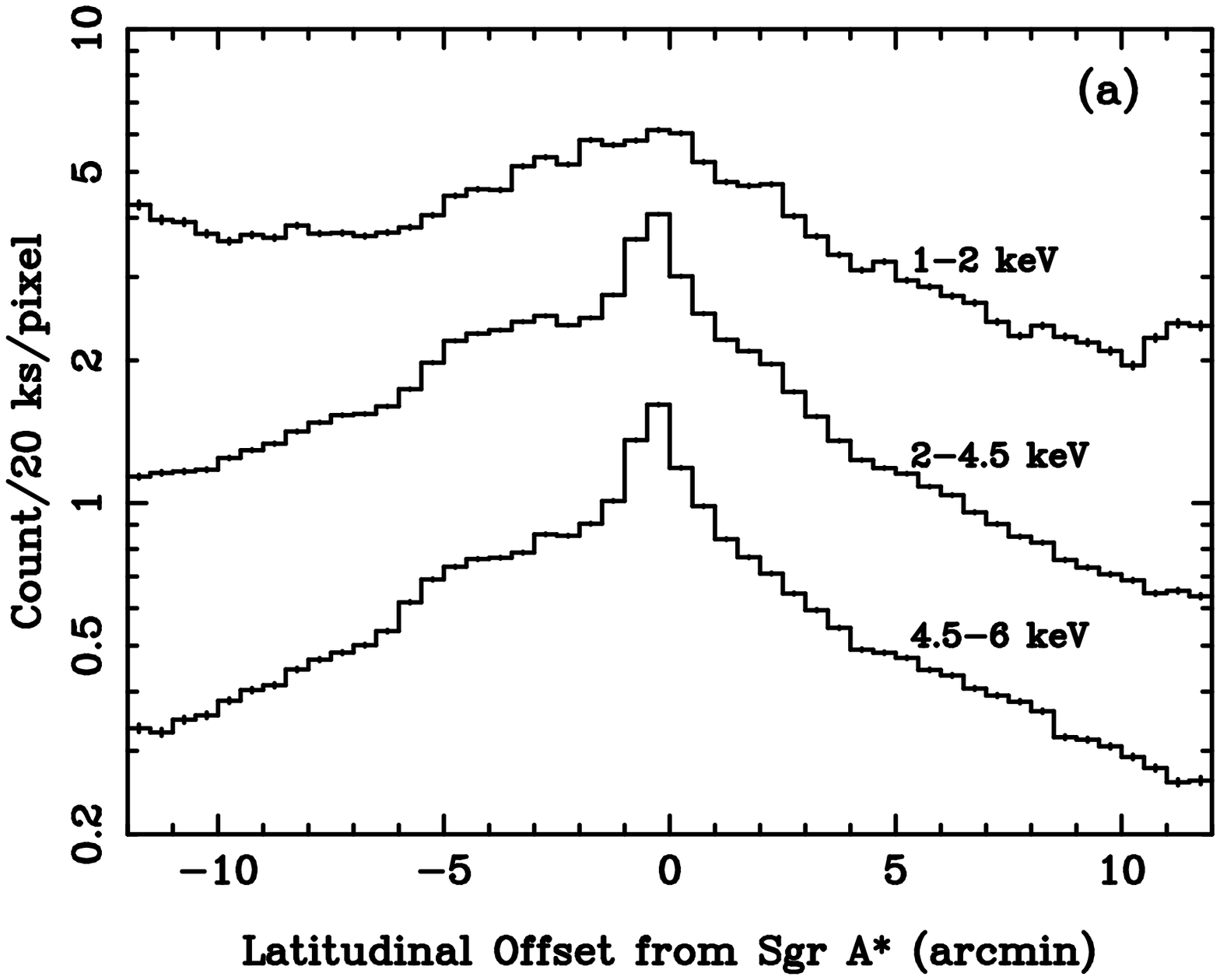}&
\includegraphics[trim=-3mm 0mm 0mm 0mm,clip,width=55mm, angle=270]{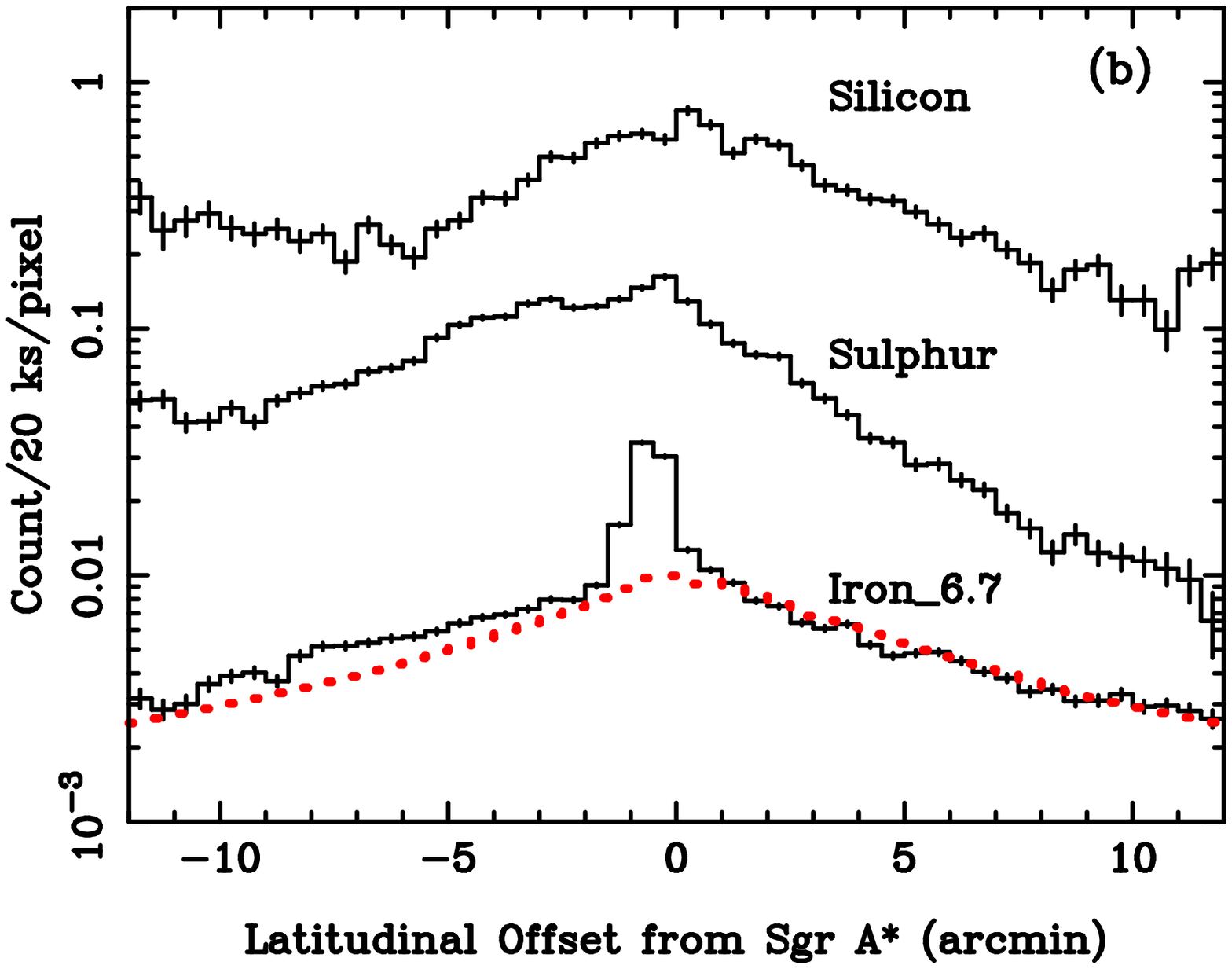}\\
\includegraphics[width=55mm,angle=270]{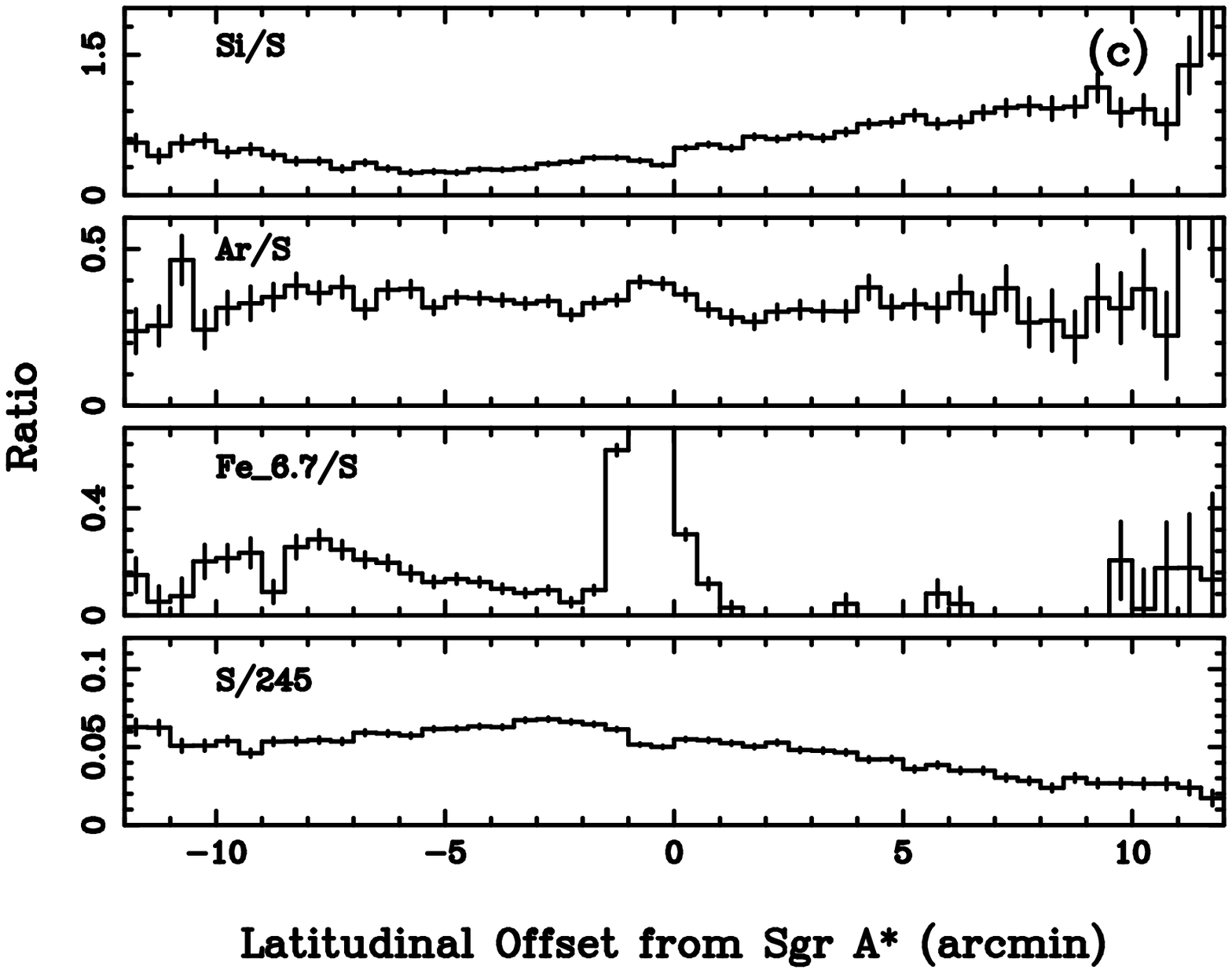}&
\includegraphics[width=55mm,angle=270]{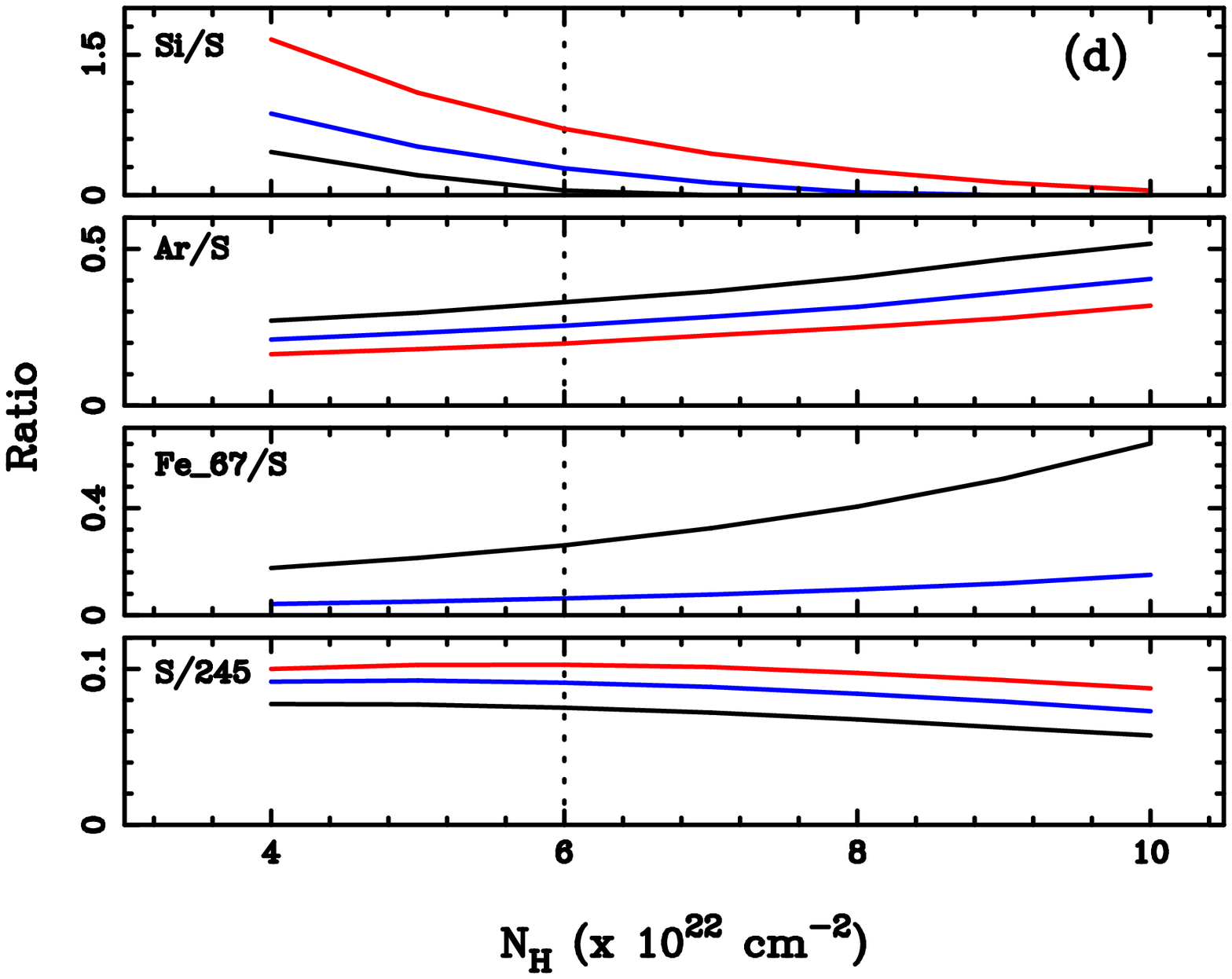}\\
\end{tabular}
\caption{Spatial cuts along the direction  of Galactic latitude for
a $12\arcminn \times 12 \arcmin$ region  centred on  Sgr A*. 
\textit{Upper-left (a):}  The distribution  of
soft continuum emission in  the three ``broad'' bands encompassing the
1--6 keV  energy range.   For clarity, the  1--2 keV  distribution has
been  shifted  upwards by  a factor  of  10. \textit{Upper-right (b):}  The 
distribution  of  the  emission in  three  narrow  bands
corresponding  to the He-like  Si,  He-like S  and He-like
Fe  K-shell  lines. Again  for
clarity, the Si-line (Fe-line) distribution has been shifted upwards
(downwards)  by a  factor of  10. The  dotted red  line  indicates the
unresolved-source  contribution to the  He-like Fe line -- see text.
\textit{Lower-left (c):}  Various
count-rate  ratios as follows:  Si/S; Ar/S,  (He-like) Fe/S  (with the
unresolved source contribution to  the iron emission subtracted). 
Also shown is the fractional contribution of the S line to
the  2--4.5  broad  band  emission.  \textit{Lower-right (d):}  The
predicted ratios  as a function  of the assumed  line-of-sight $N_{\rm{H}}$
for thermal plasmas at temperatures, $kT$, of 0.8 keV (red line), 
1.0 keV (blue line) and 1.2 keV (black line). The vertical dashed line
is drawn at $N_{\rm{H}} = 6 \times 10^{22} \rm~cm^{-2}$, the nominal 
line-of-sight column density to the GC.}
\label{fig:cuts}
\end{figure*}

A number of  distinct spatially extended  structures are  apparent in
Fig. \ref{fig:245}, on scales ranging from sub-parsec up to tens of parsec
({\it e.g.,} \citealt{baganoff03};
\citealt{morris03}; \citealt{muno03}, \citeyear{muno04}; \citealt{park04}).
The brightest feature 
near the centre of the image is the Sgr A East SNR (\citealt{maeda02};
\citealt{sakano04}; \citealt{koyama07a}).  Beyond this,
the soft X-ray emission within a  radius of about 6 arcmin of Sgr A* appears
to have  a clumpy ``bipolar''  morphology with a long  axis orientated
roughly perpendicular to the Galactic Plane (Fig. \ref{fig:245}, region A) -- see \S\ref{sec:bipolar}.
The  soft  X-ray  emission  is  also  significantly  enhanced  to  the
NE of  Sgr A* (Fig. \ref{fig:245}, region B). Here, the spatial morphology is  rather  complex  with
evidence  for  several  narrow  filamentary features. The surface
brightness peaks at the location of a discrete source (G0.13-0.11)
previous  described as a potential PWN (\citealt{wang02}).
We  will discuss  the nature  of  the emission  in this region  in
\S\ref{sec:NE}.  The  final distinct large-scale  structure evident in
Fig. \ref{fig:245}  lies to the south of  Sgr A*, in the  form of an
elongated  loop  feature  which  is  apparent  in  the  lower  surface
brightness emission (Fig. \ref{fig:245}, region C). \citet{mori09} have interpreted this feature as a
superbubble    --    we    will    consider   this    possibility    in
\S\ref{sec:sb}.

\subsection{Latitudinal cuts}
\label{sec:cuts}

We have explored  the general  properties of the  GC soft X-ray  emission by
taking cuts in Galactic latitude through several of the broad-band images
and also through a number of the 
narrow-band spectral-line images. The  region considered is that shown
in Fig. \ref{fig:245} (as the red  box) which is centred on Sgr A* and
has an extent of $\pm 12$ arcmin in both the Galactic latitude and longitude
directions.  The  resulting cuts  represent  count-rate averages  over
longitude,   sampled   at   a   30 arcsec spacing   in   latitude. 
Throughout this section we reference the latitude
offset to the plane that passes through Sgr A*.

The  cuts through  the 2--4.5  keV and  4.5--6 keV  images peak  at the
position of Sgr A East SNR (Fig. \ref{fig:cuts}a). The surface brightness in
these  bands falls  off sharply  with the latitude  offset, albeit  with a
distinct ``pedestal'' of emission at offset angles between $-1$
and $-6$ arcmin.
In  the 1--2 keV band, Sgr A
East is much less prominent  and the  emission  distribution is  somewhat
flatter than that seen at  higher energies.  This suggests that the 
contribution  from a region (or regions) located in front of the GC
(where $N_{\rm{H}}$ is somewhat reduced) may be significant, particularly
outside of the central $\sim 5$ arcmin radius zone.

Similar  patterns
can be traced in the cuts  deriving  from  the  Si,  S  and  Fe67
images (Fig. \ref{fig:cuts}b).  The Sgr  A East SNR
is a very bright  source in the Fe67 image but makes
little   contribution  in  the lower energy lines (\citealt{maeda02};
\citealt{sakano04}; \citealt{koyama07a}).   The
distribution of the S-line emission falls off steeply towards +ve offset, but
less so towards -ve offset. In contrast, the Si-line emission generally has a
flatter distribution,  particularly at -ve offset.  The  fact that
both emission lines are bright across the full field confirms that
soft-thermal emission pervades the region.
Of the three lines, the  Fe67 distribution is most symmetric with
respect to  the latitudinal offset from  Sgr A*.  As noted  in Paper I,
much  of this Fe-line flux  can  be accounted  for in  terms of  the
integrated   emission  of   unresolved  low-luminosity   sources.   In
the figure  we  show the  estimated
contribution to the Fe67 emission of the unresolved sources,
as determined by the source model reported in Paper I. Clearly, this model
accounts for virtually all of the observed emission at +ve offset, 
although a modest excess is evident at -ve offset
($-2$ to $-10$ arcmin).  In Paper
I  we interpreted  this excess He-like iron emission
as  arising from  thermal plasma  with a
temperature extending up to 1.5  keV, {\it i.e.,} a somewhat hotter
temperature component than is the norm for the GC region.
We will return to this point shortly.

Fig. \ref{fig:cuts}c compares the signal
recorded in  several bands to that  observed in the case of the S line,
with Fig. \ref{fig:cuts}d providing  a  calibration of  the
measured ratios in terms of various  \texttt{vapec} thermal  models.  
In this  latter  plot, the x-axis has  $N_{\rm{H}}$
ranging from $(4 -10) \times 10^{22}$ cm$^{-2}$ and the three curves
corresponding  to plasma temperatures, $kT$, of 0.8, 
1.0 and 1.2  keV. All  the models assume solar abundances.

The Si/S ratio reaches a minimum value of 0.24 at -ve offset,
consistent with a 1-keV thermal component absorbed by
$N_{\rm{H}} = 6 \times 10^{22} \rm~cm^{-2}$ (the nominal GC column density 
assumed by \citealt{muno04}). The rise in the Si/S ratio
towards +ve offset can be matched by a reduction in the net
line-of-sight column to $N_{\rm{H}} \sim 4 \times 10^{22} \rm~cm^{-2}$.
This might be explained either in terms of a reduction in
the column density affecting the bulk of GC plasma or due
to an increasing contribution from the foreground emission as the
GC component falls off with latitude.
The Ar/S ratio remains reasonably constant across the whole field
at a value of $\sim$0.3; this again matches a thermal component
with a temperature of roughly 1 keV with the nominal GC $N_{\rm{H}}$.
The Fe67 excess peaks around $-8$ arcmin, with a Fe67/S ratio 
commensurate with a (somewhat enhanced) plasma temperature of about
1.2 keV. Alternatively a blend of temperatures from, say,  1--1.5 keV
could also explain the observed He-like iron excess\footnote{
The centroid of the iron-line complex pertaining to
a plasma with a temperature in the range 1--1.5 keV 
peaks below 6.7 keV, indicative of a range of ionization states
below that of He-like iron. Nevertheless
a significant signal will still be recorded in
our narrow-band He-like Fe channel. There will also be some
spillover into the 6.4 keV narrow-band. which may explain the
modest 6.4-keV excess at -ve offset evident in Fig. 4 of
Paper I.}.

Fig. \ref{fig:cuts}c and  Fig. \ref{fig:cuts}d also show the S/245 ratio, 
that is the contribution of the S line to the
2--4.5 keV broad-band signal.  The measured ratio peaks at a value of
$\sim 0.07$ at -ve offset and then declines at +ve offsets
to a value of $\sim 0.03$. In contrast, the predicted fraction
for our fiducial thermal model ($kT = 1$ keV,
$N_{\rm{H}} = 6 \times 10^{22} \rm~cm^{-2}$) is 0.095. This discrepancy is
readily explained if the contribution
of unresolved sources to the measured 2--4.5 keV broad-band
signal is roughly 30 per cent in the region where the S/245 ratio
is at its peak value, ranging up to a 70 per cent contribution 
at the northern boundary of the region studied.

\section{A bipolar outflow from Sgr A*?}
\label{sec:bipolar}

\subsection{Images of the central region}
\label{sec:imagesbi}

\begin{figure*}
\centering
\begin{tabular}{cc}
\includegraphics[width=63mm,angle=0]{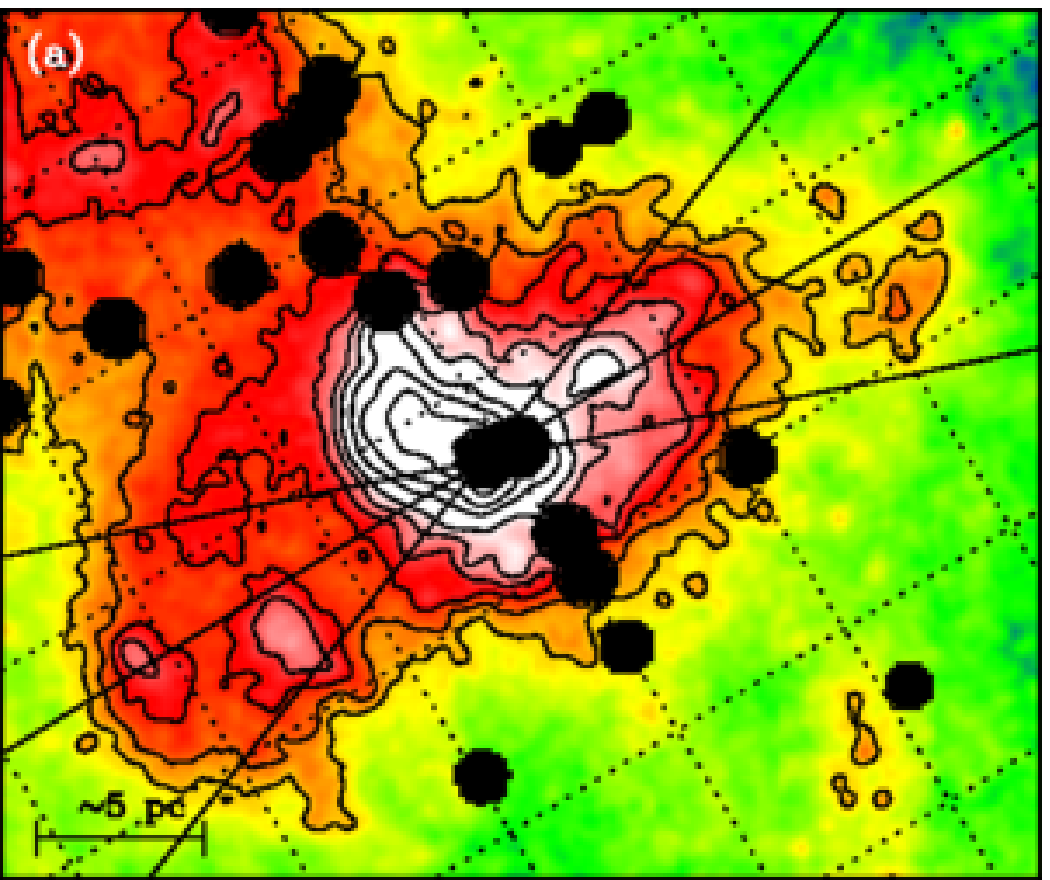}&
\includegraphics[width=63mm,angle=0]{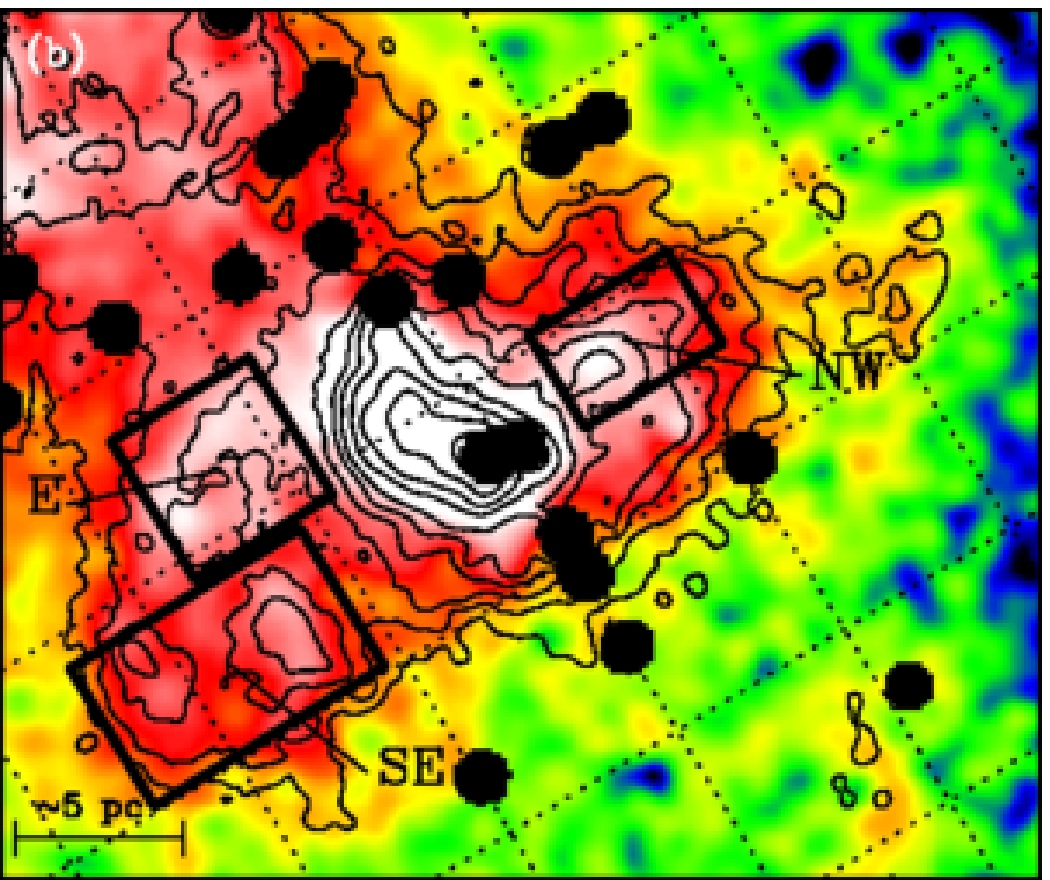}\\
\includegraphics[width=63mm,angle=0]{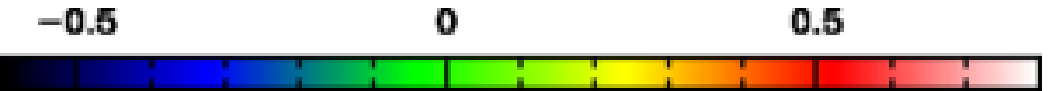}&
\includegraphics[width=63mm,angle=0]{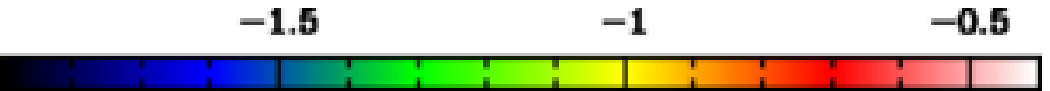}\\
\includegraphics[width=63mm,angle=0]{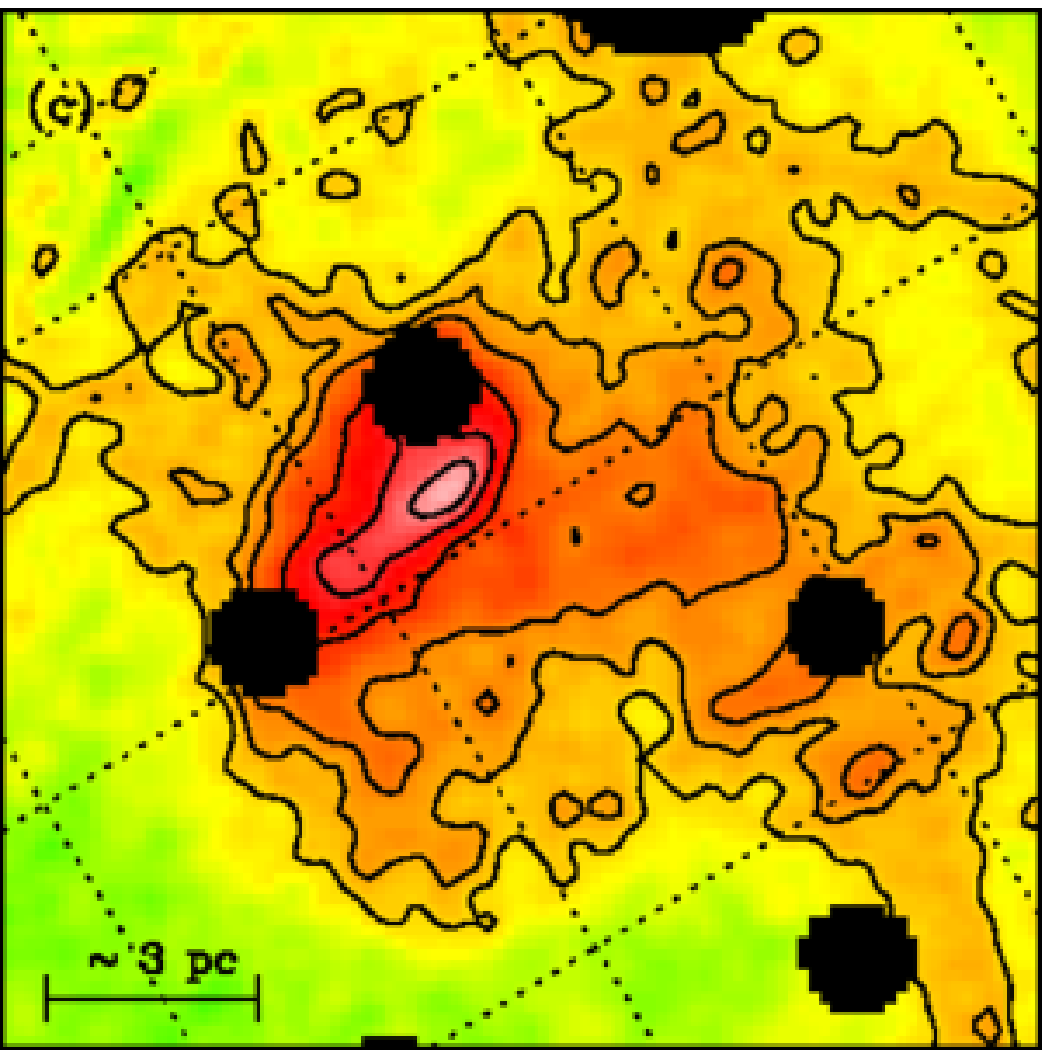}&
\includegraphics[width=63mm,angle=0]{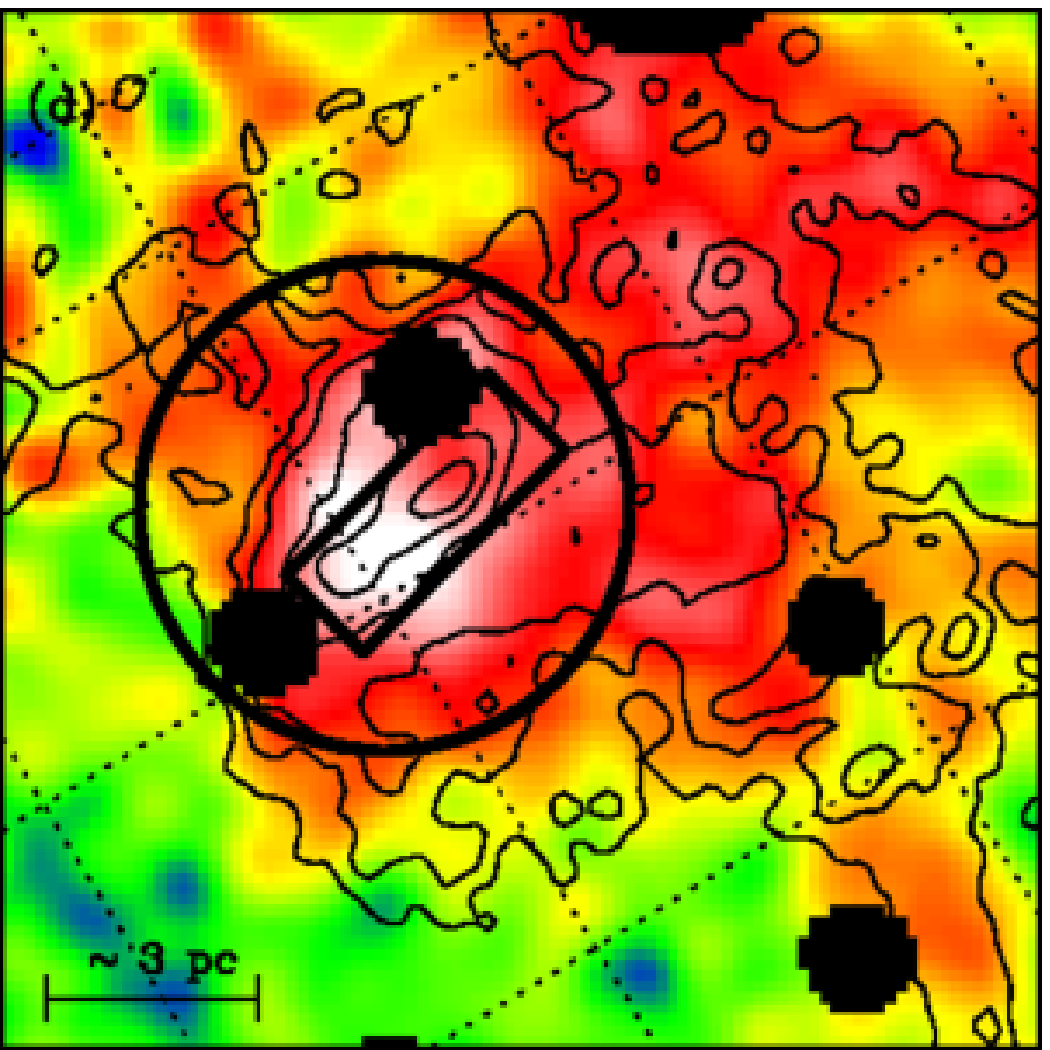}\\
\includegraphics[width=63mm,angle=0]{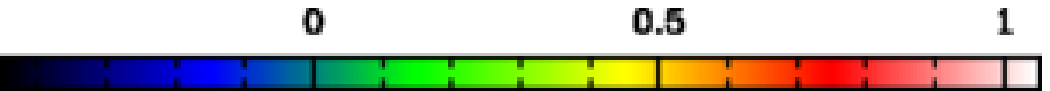}&
\includegraphics[width=63mm,angle=0]{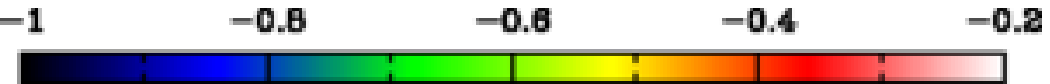}\\
\includegraphics[width=61mm,angle=0]{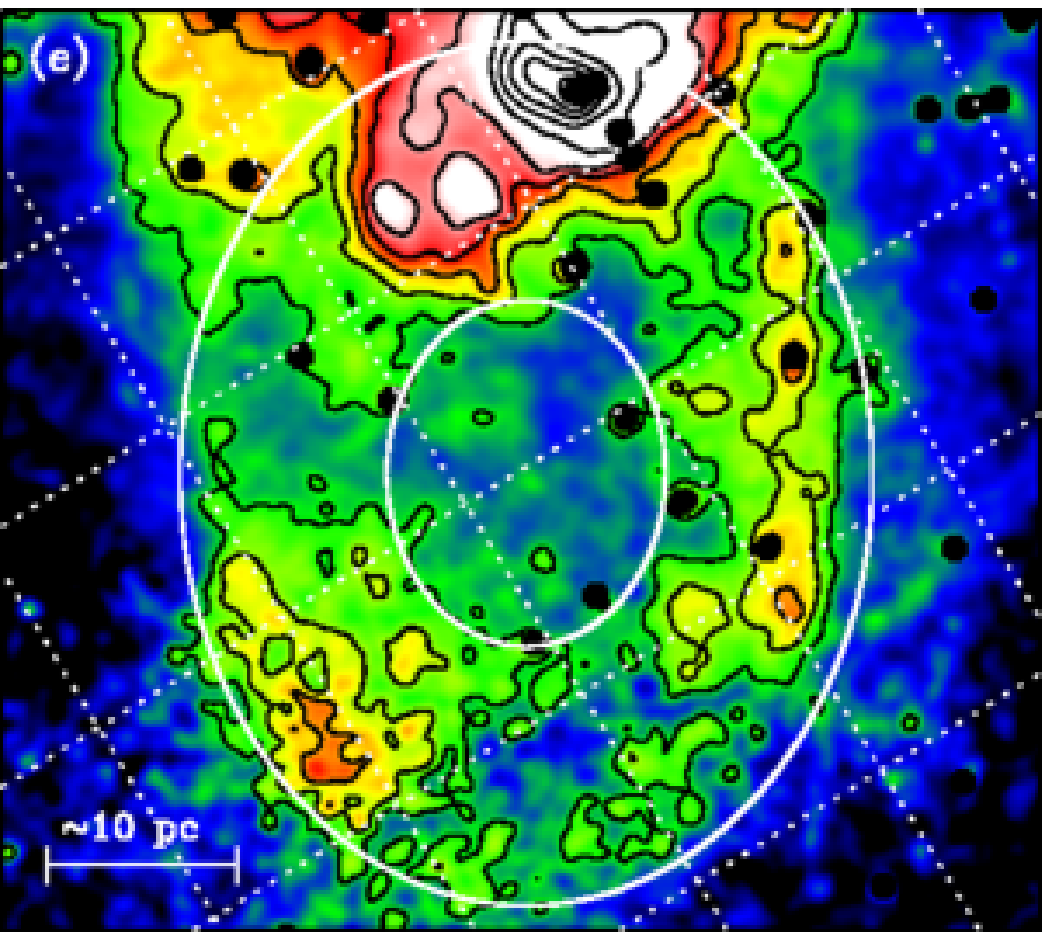}&
\includegraphics[width=61mm,angle=0]{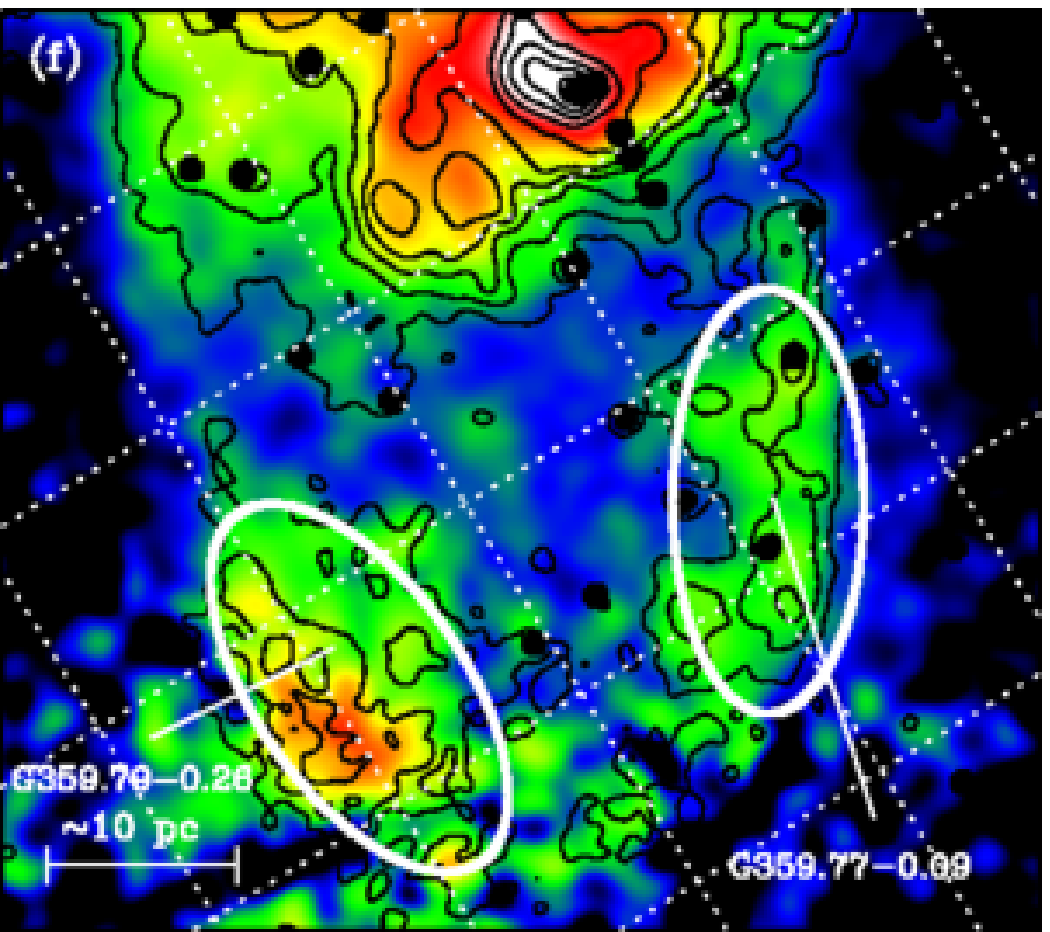}\\
\includegraphics[width=63mm,angle=0]{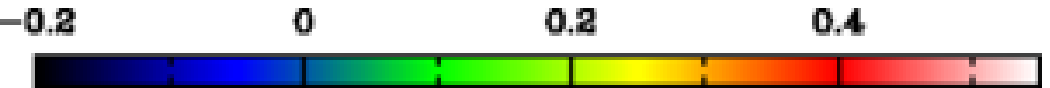}&
\includegraphics[width=63mm,angle=0]{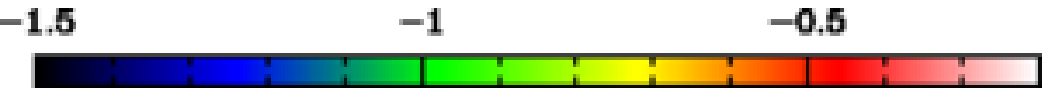}\\
\end{tabular}
\caption{ {\bf (a)} 2--4.5 keV image of the central $12.9\arcminn \times 10.7\arcmin$ region.  The solid diagonal lines are orientated at -20, 0 and +20 deg relative to a line of constant longitude passing through Sgr A*.  {\bf (b)}  Combined S and Ar (S$+$Ar) image of the same region as in (a), with contours derived from the 2--4.5 keV image. The rectangles labelled NW, SE and E indicate the regions from which spectra were extracted. {\bf (c)} 2--4.5 keV image of the north-east enhancement region covering a $6.3\arcminn \times 6.3\arcmin$~ field. {\bf (d)} The S$+$Ar image of the same region as in (c), with contours derived from the 2--4.5 keV image. The  rectangle  represents the spectral extraction region for the  PWN  G0.13-0.11, whereas  the circle defines the spectral extraction extent for the surrounding nebulosity. {\bf (e)} 2--4.5 keV image of the superbubble region covering a $23\arcminn \times 21\arcmin$ field. The two ellipses define the possible extent of the superbubble structure -- see text. {\bf (f)} The S$+$Ar image of the same region as in (e), with contours derived from the 2--4.5 keV image. The two  small ellipses encompass the features designated as G359.77-0.09 and G359.79-0.26 for which spectra were extracted. In all the images the intensity scaling is logarithmic, with the colour bar shown at the bottom  of each panel having units of log$_{10}$  (count/20 ks/pixel). The diagonal dotted lines represent a Galactic coordinate grid with 2.4 arcmin spacing in (a) -- (d) and 4.8 arcmin spacing in (e) -- (f). The continuum images have been spatially smoothed using a Gaussian function of width $\sigma$ = 4 arcsec, whereas for the S$+$Ar images $\sigma$ = 10 arcsec (except for the superbubble region where the image smoothing parameters were $\sigma$ = 10 arcsec and $\sigma$ = 20 arcsec respectively). The filled black circles correspond to sources excised by the spatial mask.}
\label{fig:images}
\end{figure*}

\begin{figure}
\centering
\begin{tabular}{c}
\includegraphics[width=60mm,angle=270]{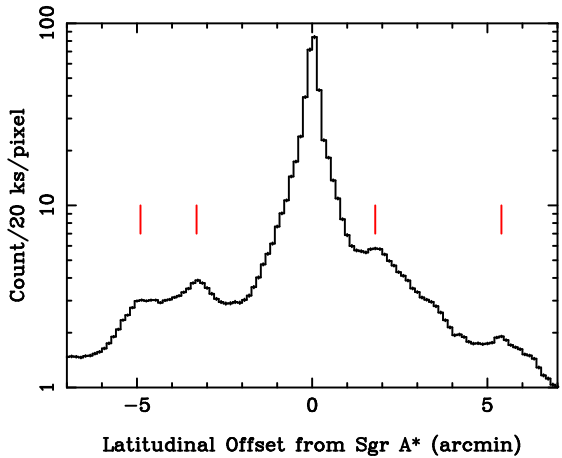}
\end{tabular}
\caption{The radial distribution of the 2--4.5 keV emission
within the bounds of the conical regions shown in
Fig. \ref{fig:images}a. The vertical lines
drawn at offset angles of $-4.9$, $-3.3$, $+1.8$ and
$+5.4$ arcmin mark the positions of the two SE clumps, the NW ridge and
the isolated NW cloud respectively.
}
\label{fig:cone}
\end{figure}

Fig.  \ref{fig:images}a shows a  zoomed-in version of
the   2--4.5  keV   image   covering  a central 
$12.9\arcminn \times 10.7\arcmin$ (30 pc $\times$ 25 pc) region. As noted 
previously the brightest extended object in the field is the Sgr A
East SNR. On  a  somewhat larger  scale,  the  emission distribution
appears to be elongated along an axis perpendicular to the Galactic 
Plane. The twin conical regions defined by the diagonal lines in the figure
(with an apex at the location of Sgr A* and an opening half-angle of
20\degn) encompass several of the bright features
which, in effect, delineate this elongated structure. This includes  
a ``ridge'' of high surface brightness extending 
approximately 2 arcmin (4.6 pc) to the north-west and two obvious bright clumps 
located between $3-6$ arcmin (7--14 pc) to the south-east. A further isolated 
lower-surface brightness ``cloud'' is also present in the north-west
sector.  These features are also evident in the radial profile of the
emission shown in Fig. \ref{fig:cone}.
This figure highlights the presence of an asymmetry in the emission
distribution, namely that the radial profile to the north-west is markedly
steeper than that to the south-east.  As previously reported by  
\citet{sidoli99},
an elongated structure similar to that seen in X-rays is also apparent
in the synchrotron halo of Sgr A, as revealed in 90 cm radio observations
(\citealt{larosa00}; \citealt{nord04}). Finally we note that a further
arc-like feature is evident in the {\it XMM-Newton} images. 
This structure extends to the south-east,
from its starting point at the eastern boundary of the Sgr A East SNR, 
over a length scale of at least 2.5 arcmin (6 pc).

Fig.  \ref{fig:images}b shows for comparison
the S+Ar line image obtained by co-adding the individual S and
Ar images of this central region. The same features, as
noted above, are bright in this line image, demonstrating
their soft-thermal nature. We have investigated the spectra
of the three regions defined by black rectangles in the line
image, labelled as the NW, SE, and E regions; these encompass,
respectively, the north-west ridge, the two south-east clumps and
the arc-like feature.

\subsection{Spectral constraints}
\label{sec:specbi}

\begin{figure*}
\centering
\begin{tabular}{cc}
\includegraphics[width=60mm,angle=270]{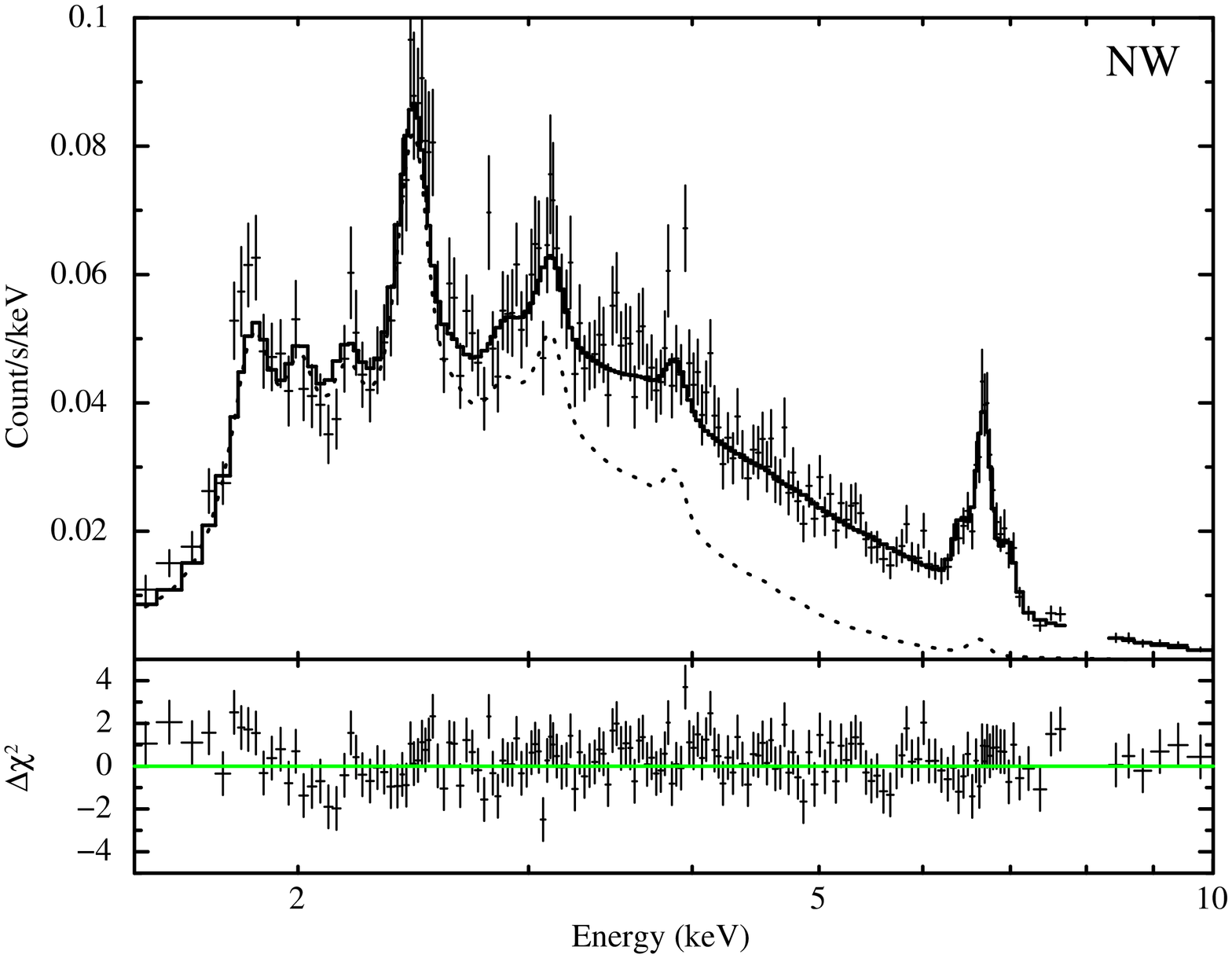}&
\includegraphics[width=60mm,angle=270]{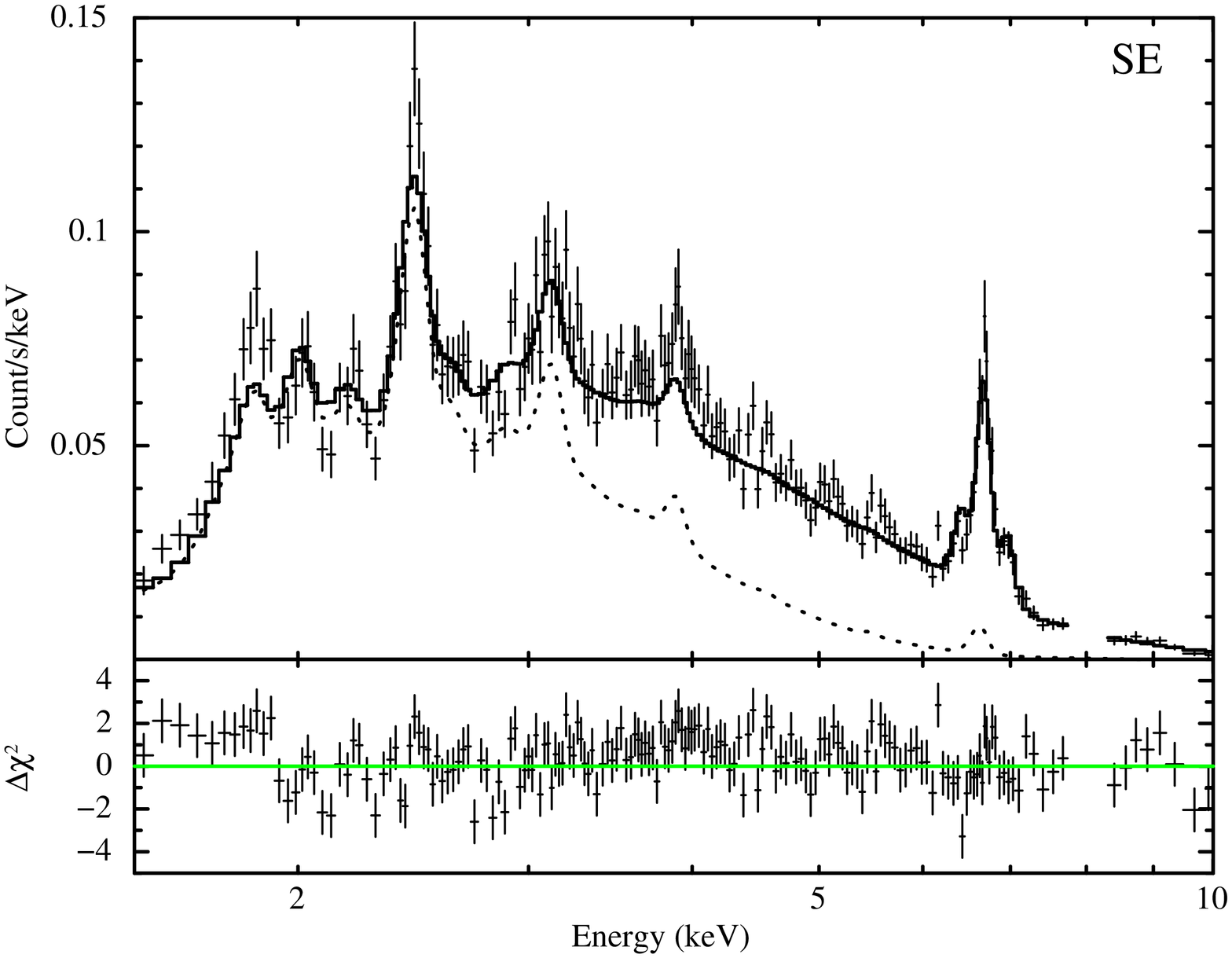}\\
\includegraphics[width=60mm,angle=270]{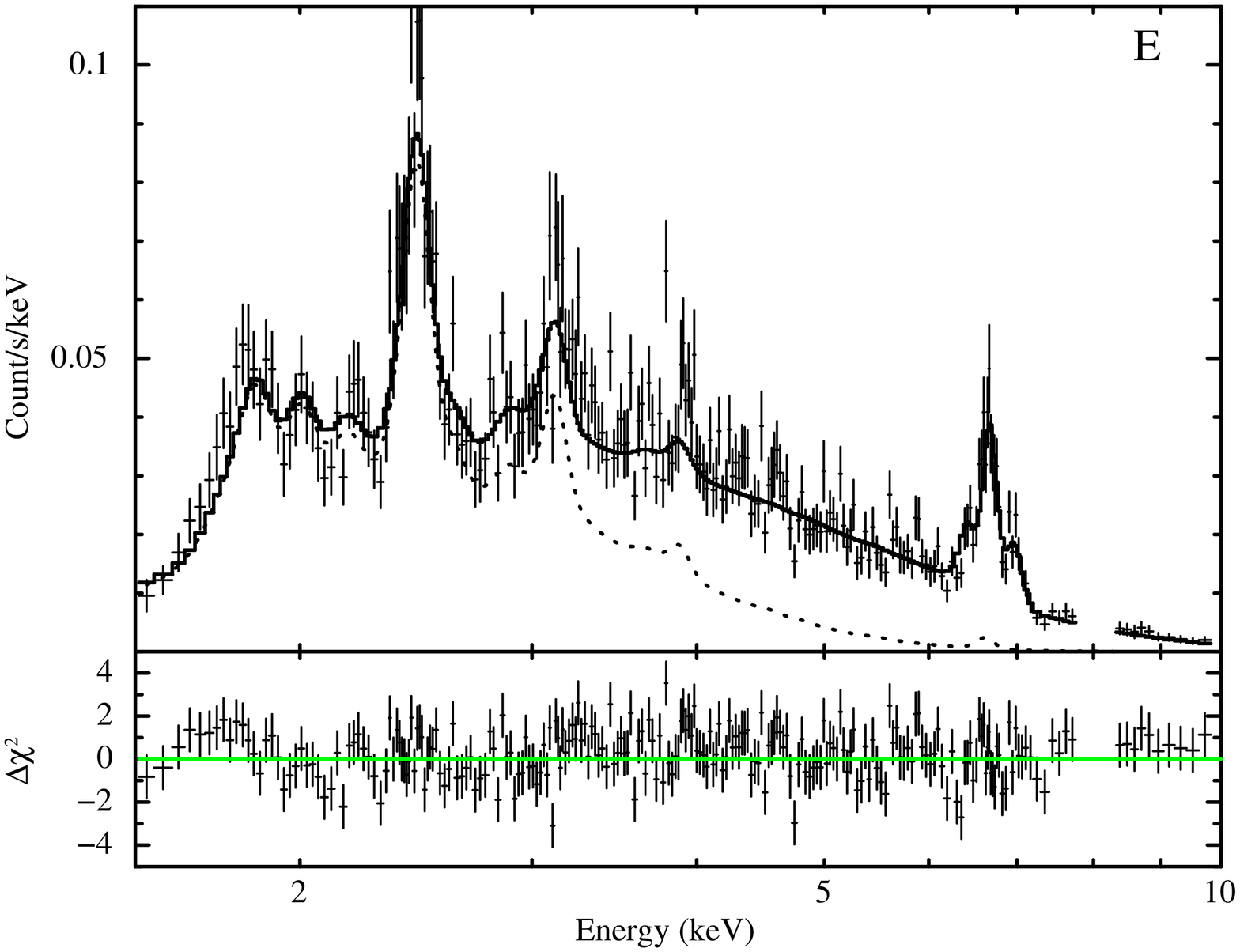}&
\includegraphics[width=60mm,angle=270]{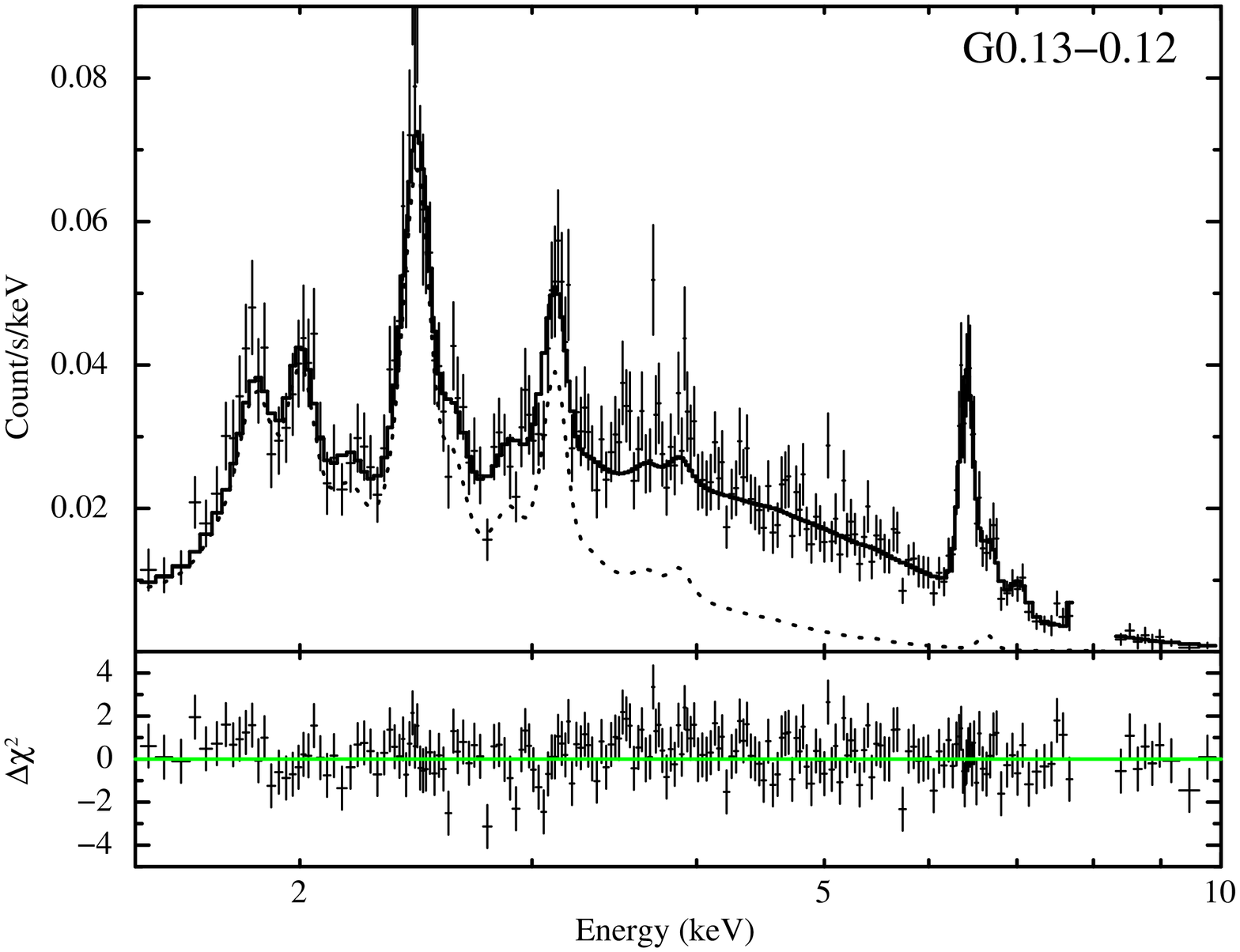}\\
\includegraphics[width=60mm,angle=270]{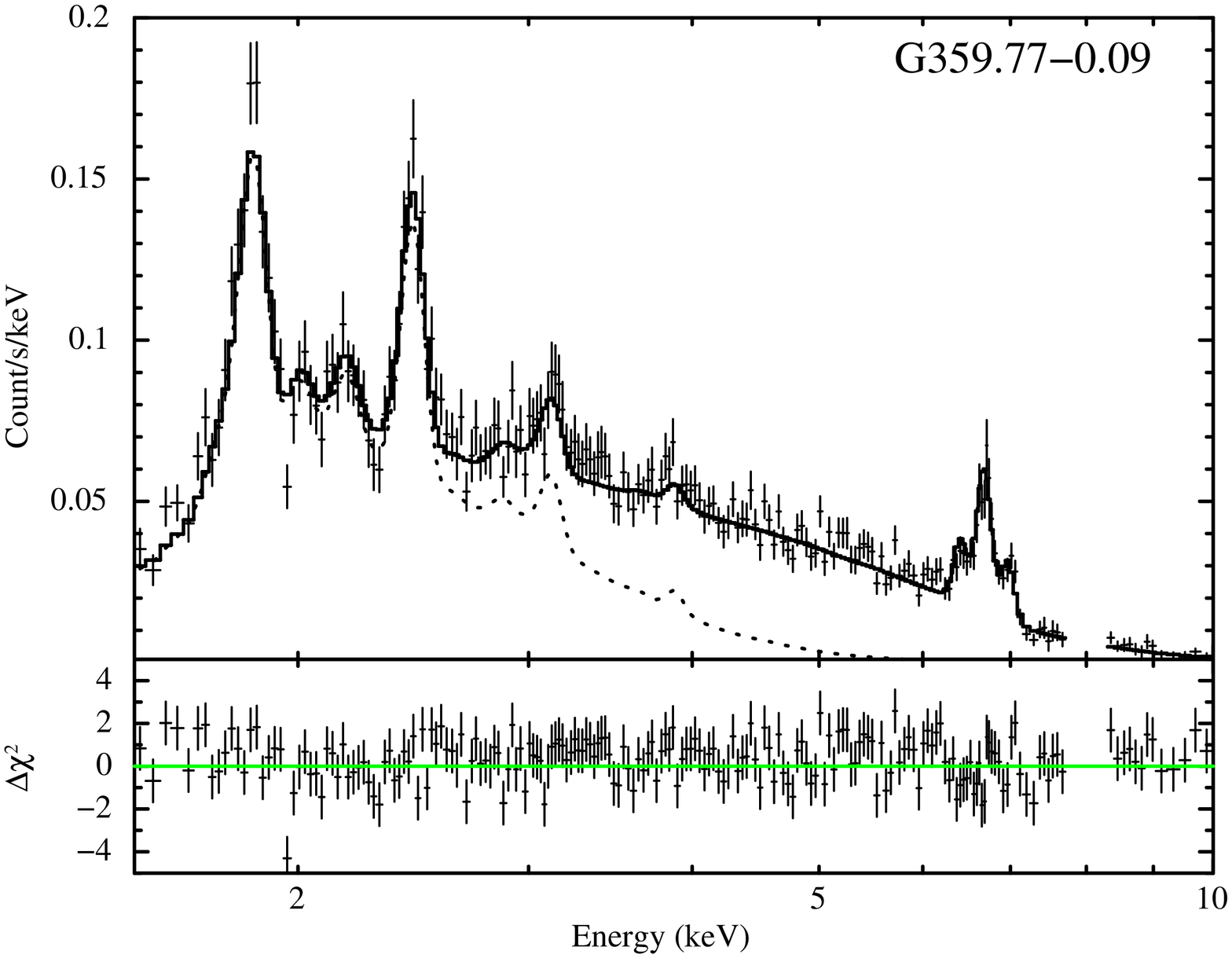}&
\includegraphics[width=60mm,angle=270]{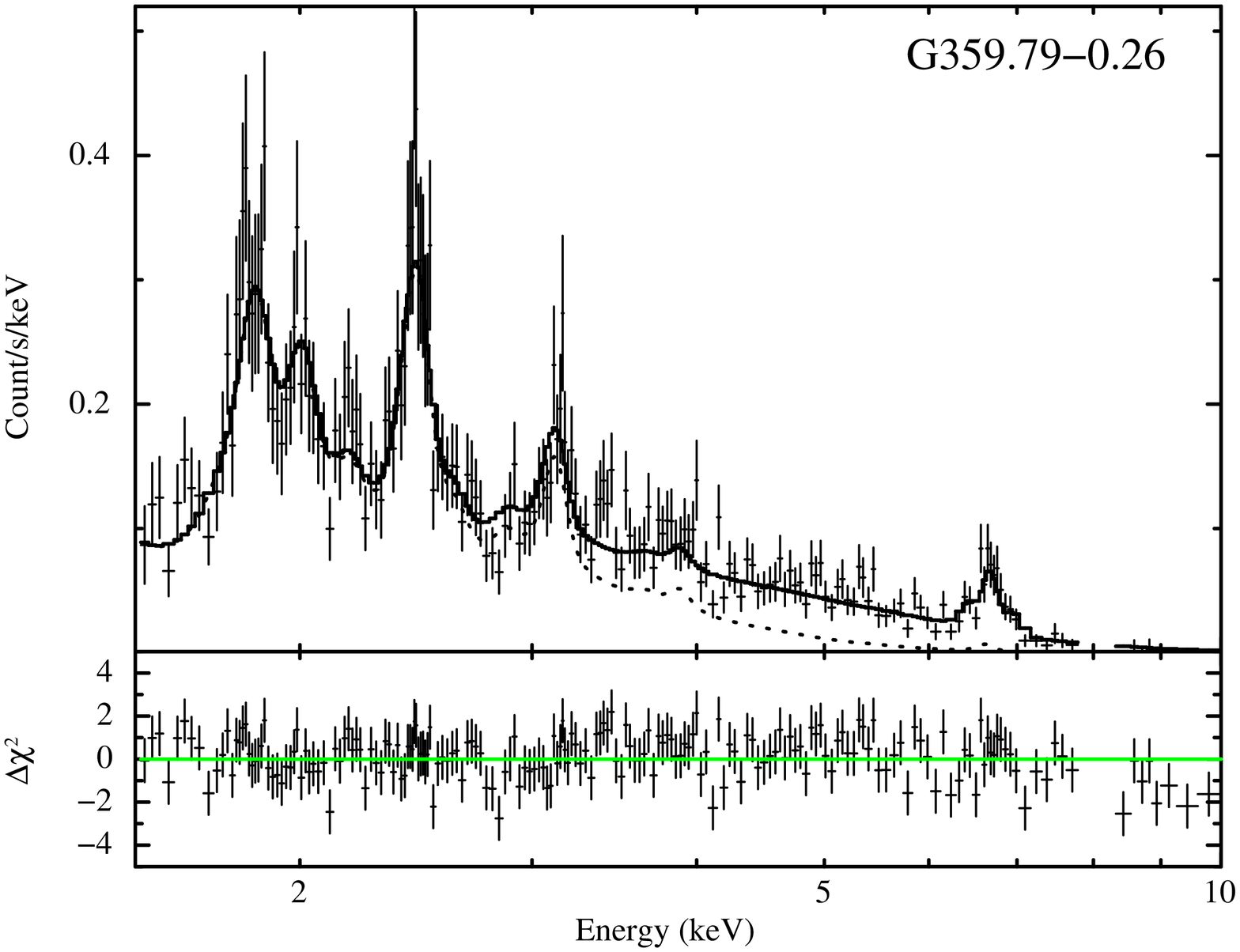}\\

\end{tabular}
\caption{Best-fitting pn spectra of the six regions. In each case
the spectral data and best-fitting composite model (solid black line)
are shown together with the contribution of the soft-thermal
emission to the fit (dotted line). The residuals to the best-fit model
are also shown.  From left to right, top to bottom the panels
show the results for the NW, SE, E, G0.13-0.12, G359.77-0.09 and G359.79-0.26 regions.}
\label{fig:spectra}
\end{figure*}

The spectra extracted from the three regions identified above
can be modelled assuming just two emission components,
namely the emission from unresolved sources and soft-thermal emission
(see \S\ref{sec:spectralx}).
The resulting best-fitting parameters pertaining to the thermal
plasma emission are reported in the first three
columns of Table  \ref{tab:bestfit}. Fig. \ref{fig:spectra}
shows the corresponding best-fitting spectra and residuals.
The two-component model provides a reasonable fit to the data in all
three regions, although systematic trends in the residuals, particularly below 2 keV,
emphasise the approximate nature of our single-column, single-temperature,
thermal model.
The plasma temperature characterising  the three
regions is not too far from the 1-keV fiducial discussed in \S\ref{sec:cuts},
although the temperature measured in the SE region is marginally hotter
than that in the other two regions (by $\sim$ 0.2 keV).
In Fig. \ref{fig:spectra}, this higher temperature is most evident
in terms of the enhanced line emission near 6.7 keV (comparing the 
modelled spectrum of the SE region with that of the NW and E regions)
due to the Fe-K emission of highly stripped iron ions up to
helium-like states.
The absorption column density measured for the NW region is 
somewhat higher than that determined for the  SE and E  regions
and also higher than the nominal column density of
$6 \times 10^{22} \rm~cm^{-2}$ reported for the GC.  One interpretation
of this additional absorption might be that the north-east ridge lies behind the
plane of Sgr A*.  If this were the case then it would imply
that the northern component of the elongated structure is
tipped both away from us and away from the normal to the Galactic Plane
-- see below.
Of course, the additional absorption may simply be a manifestation of the
complex distribution of molecular gas in the immediate foreground of
the Sgr A region. A further result from the spectral fitting is that the 
relative abundances determined for Si, S and Ar are near
to solar values for all three regions.

\begin{table*}
\begin{center}
\caption{The best-fitting spectral parameters for the thermal emission
seen in the NW, SE, E, G0.13-0.12, G359.77-0.09 and G359.79-0.26 regions.}
\begin{tabular}{l c c c c c c}
\hline
&\multicolumn{3}{c}{Bipolar outflow}&SNR &\multicolumn{2}{c}{Superbubble}\\
& NW & SE & E & G0.13-0.12 & G359.77-0.09 &G359.79-0.26 \\
\hline
Area (arcmin$^{2}$)&1 $\times$ 0.7&1.7 $\times$ 1&1 $\times$ 1&$\pi$1.5$^{2}$&4.8 $\times$ 2.2&4.8 $\times$ 2.4\\
\hline
$N_{\rm{H}}$ ($\times10^{22}$ cm$^{-2}$)&8.0$^{+0.6}_{-0.5}$&6.3$^{+0.3}_{-0.3}$&6.3$^{+0.4}_{-0.4}$&5.6$^{+0.4}_{-0.4}$&5.9$^{+0.3}_{-0.2}$&4.4$^{+0.5}_{-0.4}$\\
$kT$ (keV)&0.91$^{+0.06}_{-0.05}$&1.14$^{+0.06}_{-0.05}$&0.96$^{+0.07}_{-0.06}$&1.1$^{+0.1}_{-0.1}$&0.73$^{+0.04}_{-0.04}$&1.02$^{+0.15}_{-0.08}$\\
$Z_{\rm{Si}}$ ($\Zsunn$)&1.0$^{+0.4}_{-0.3}$&0.6$^{+0.2}_{-0.2}$&0.7$^{+0.3}_{-0.2}$&1.4$^{+0.5}_{-0.3}$&0.6$^{+0.1}_{-0.1}$&1.1$^{+0.4}_{-0.3}$\\
$Z_{\rm{S}}$ ($\Zsunn$)&0.7$^{+0.1}_{-0.1}$&0.7$^{+0.1}_{-0.1}$&1.2$^{+0.2}_{-0.2}$&2.0$^{+0.8}_{-0.5}$&0.7$^{+0.1}_{-0.1}$&1.3$^{+0.3}_{-0.3}$\\
$Z_{\rm{Ar}}$ ($\Zsunn$)&0.6$^{+0.3}_{-0.3}$&0.9$^{+0.3}_{-0.3}$&1.5$^{+0.5}_{-0.5}$&3.4$^{+1.7}_{-1.0}$&1.0$^{+0.4}_{-0.3}$&2.3$^{+1.0}_{-0.9}$\\
$Norm.$ ($\times 10^{-2}$) &2.1$^{+0.4}_{-0.4}$&1.6$^{+0.2}_{-0.2}$&1.2$^{+0.2}_{-0.3}$&1.0$^{+0.7}_{-0.9}$&4.9$^{+1.1}_{-1.0}$&2.8$^{+1.0}_{-0.9}$\\
$F_{2-10}$ (10$^{-12}$ erg s$^{-1}$ cm$^{-2}$)$^{a}$&0.7$^{+0.1}_{-0.1}$&1.1$^{+0.1}_{-0.1}$&0.6$^{+0.1}_{-0.1}$&1.1$^{+0.3}_{-0.3}$&1.2$^{+0.3}_{-0.2}$&2.6$^{+0.7}_{-0.8}$\\
$F_{2-10}$ (10$^{-12}$ erg s$^{-1}$ cm$^{-2}$)$^{b}$&3.2$^{+0.6}_{-0.6}$&3.7$^{+0.5}_{-0.5}$&2.1$^{+0.4}_{-0.6}$&3.1$^{+0.9}_{-0.9}$&4.4$^{+1.0}_{-0.9}$&6.5$^{+1.9}_{-2.1}$\\
\hline
$\chi^{2}$/$\nu$&533/511&862/724&547/479&543/558&804/824&208/191\\
\hline
\end{tabular}
\label{tab:bestfit}
\end{center}
\flushleft{$^{a}$The absorbed flux in the 2--10 keV band.~~~~$^{b}$The unabsorbed flux in the 2--10 keV band.}

\end{table*}

Using the results from Table \ref{tab:bestfit}, we have calculated the
physical parameters of the X-ray emitting plasma  contained within the NW,
SE  and E  regions --  see Table  \ref{tab:plasmax}.  In  each case  we
assumed that  the dimension  of the emitting  region in  the direction
along the line-of-sight was equal to  that of the smaller dimension of
the  defining  rectangle (see Table  \ref{tab:bestfit}). 
The  X-ray luminosities  are  determined from  the
unabsorbed  fluxes quoted  in Table  \ref{tab:bestfit}.  A number of
the physical parameters depend on the volume filling factor of the emitting
plasma ($f$) as indicated. The  cooling
time  ($t_{\rm{cool}}$=$E_{\rm{th}}$/$L_{\rm{bol}}$)   was  determined
assuming  an $L_{\rm{bol}}$ to $L_{\rm{X}}$ ratio of  20, which  has an
associated uncertainty of about 50 per cent \citep{muno04}.

The  parameter values  pertaining to the hot plasmas in the  three
regions  are fairly  similar.  The inferred  electron densities are in
the range 4--10 cm$^{-3}$ (assuming $f$ is of order  unity), 
which is higher than is typical of the GC in general ({\it e.g.,} \citealt{muno04};
\citealt{koyama96}).   The total mass of hot plasma contained within 
each region is roughly $1-2~ \Msunn$.  Similarly the radiative cooling
timescales are all in the range $(5-10) \times 10^{5}$ yr. 

\begin{table*}
\begin{center}
\caption{Physical parameters of the soft thermal emission associated with the various regions.}
\begin{tabular}{l c c c c c c}
\hline
 &\multicolumn{3}{c}{Bipolar outflow}&SNR &\multicolumn{2}{c}{Superbubble}\\
 & NW & SE & E & G0.13-0.12 & G359.77-0.09 & G359.79-0.26\\
\hline
Volume, $V$ (cm$^{3}$)&$1.6 \times 10^{56}$&$5.5 \times 10^{56}$&$3.3 \times 10^{56}$&$4.8 \times 10^{57}$&$3.2 \times 10^{58}$&$3.7 \times 10^{58}$\\
Temperature, $T$ (K)&$1.1\times10^{7}$&$1.3\times10^{7}$&$1.1\times10^{7}$&$1.3\times10^{7}$&$8.5\times10^{6}$&$1.2\times10^{7}$\\
Emission integral, $EI$ (cm$^{-3}$)&1.5$\times$10$^{58}$&$1.2\times10^{58}$&8.7$\times$10$^{57}$&$7.2\times10^{57}$&3.6$\times$10$^{58}$&$2.0\times10^{58}$\\
2--10 keV luminosity, $L_{\rm{X}}$ (erg s$^{-1}$)&$2.3\times10^{34}$&$2.7\times10^{34}$&$1.5\times10^{34}$&$2.2\times10^{34}$&$3.2\times10^{34}$&$4.7\times10^{34}$\\
Electron density, $n_{\rm{e}}$ ($f^{-1/2}$ cm$^{-3}$)&9.9&4.6&5.2&1.2&1.1&0.7\\
Thermal energy, $E_{\rm{th}}$ ($f^{1/2}$ erg)&6.7$\times$10$^{48}$&$1.4\times10^{49}$&7.8$\times$10$^{48}$&$3.1\times10^{49}$&$1.2\times10^{50}$&$1.3\times10^{50}$\\
Mass, $M$ ($f^{1/2}$ $\Msun$)&$1.3$&$2.1$&$1.4$&$5.0$&$28$&$23$\\
Pressure, $P$ ($f^{-1/2}$ dyn cm$^{-2}$)&$2.9\times10^{-8}$&$1.7\times10^{-8}$&1.6$\times$10$^{-8}$&$4.3\times10^{-9}$&$2.5\times10^{-9}$&$2.3\times10^{-9}$\\
Cooling timescale, $t_{\rm{cool}}$ ($f^{1/2}$ yr)&4.6$\times$10$^{5}$&8.2$\times$10$^{5}$&8.1$\times$10$^{5}$&$2.2\times10^{6}$&5.9$\times$10$^{6}$&$4.5\times10^{6}$\\
\hline
\end{tabular}
\label{tab:plasmax}
\end{center}
\end{table*}

\subsection{Origin of the bipolar morphology}
\label{sec:originbi}

An early glimpse of the distribution of the diffuse soft X-ray emission
present in the Sgr A region was provided by \textit{BeppoSAX} observations 
\citep{sidoli99}. More recently, \textit{Chandra} has resolved
this emission into a variety of sub-components, including a number of
features which form an apparent elongated structure centred on Sgr A*
aligned perpendicular to the Galactic Plane (\citealt{baganoff03}).
These authors have suggested that the features may, collectively,
be evidence of a hot and clumpy  ``bipolar" outflow from Sgr A* -- see also 
\citet{morris03} and \citet{markoff10}. Our present {\it XMM-Newton}
observations further delineate the morphology of this structure and also
demonstrate, unequivocally, the thermal nature of the X-ray emission.

In the bipolar outflow scenario, hot plasma produced within the
central few parsecs is simultaneously driven outwards and collimated
into a channeled flow. One possibility is that the X-ray emitting
plasma is produced by shock heating when the high-velocity winds
of the many massive Wolf-Rayet and supergiant OB stars within the Central Cluster
(\citealt{genzel03}; \citealt{paumard06}; \citealt{martins07}) collide with each
other and with the surrounding ISM. \citet{rockefeller04} have shown that such a wind can fully account for the diffuse X-ray luminosity
observed within the inner 10 arcsec (0.4 pc) of the Galaxy and furthermore 
predicted a plasma temperature of $kT \approx 1.3$ keV.
As suggested by \citet{markoff10},
the resulting outflow might then be concentrated into twin ``lobes'' 
by  the 2-pc radius ring of dense molecular gas which defines
the inner extent of the Circumnuclear Disc (CND)
(\citealt{christopher05}; \citealt{oka11}). Interestingly, the orbital plane
of this dense molecular ring (position angle with respect to the Galactic plane
$p \sim 0\deg$ and inclination to our line of sight $i = 50\deg -75\degn$)
is well matched to the alignment of the bipolar X-ray structure.
One  issue with  this  hypothesis is the question of how the
sub-structure within the bipolar flow might have formed
if the wind from the Central Cluster has been
quasi-continuous over the cluster lifetime of
$6 \pm 2$ Myr \citep{paumard06}.  

Since the X-ray emitting plasma which forms the wind is likely to be
over-pressured with respect to its surroundings, adiabatic losses may dominate
over radiation cooling.  A plasma temperature of  approximately 1 keV
gives a sound speed $v_{\rm{s}}=(kT_{\rm{e}} \gamma /\mu m_{\rm{H}})^{1/2}$
$\approx500 \rm~km~s^{-1}$ (assuming energy equipartition between
the electrons and ions,  $\gamma = 5/3$ and $\mu=0.6$). On the basis of
a 20\deg half-opening angle for the bipolar structure, we can estimate
the transport velocity within the outflow to be 
$\sim500 \times \rm cosec(20\degn)~\rm~km~s^{-1}$, {\it i.e.,} $\sim1500 \rm~km~s^{-1}$.
However, this estimate will be an upper-limit if the outflow is at least partly
confined by the dense clouds located along the track of the outflow
outside of the CND region.  By way of comparison, the terminal winds for
the most massive stars in the Central Cluster range from 
$500 - 2500 \rm~km~s^{-1}$, with the average
being $\sim 1000 \rm~km~s^{-1}$ (\citealt{martins07}).

Assuming the transport velocity is $1000 \rm~km~s^{-1}$, the timescale for plasma
to reach the 6 arcmin ($14$ pc) outer extent of its flow
is $\sim 13 000$ yr, which is over an order of magnitude shorter than the
inferred radiative cooling timescale of the plasma clumps 
(Table \ref{tab:plasmax}). This is consistent with the assumption
that the adiabatic losses eventually caused the X-ray emitting plasma
to dissipate.  The kinetic energy
contained within the winds of the Central Cluster stars is at least
$8 \times 10^{38}$ erg s$^{-1}$ (summing over the 18 stars listed
in table 2 of \citealt{martins07}).  Furthermore, the modelling
of \citet{rockefeller04} shows that as much as 26 per cent of
the total wind energy within the central parsec can be converted to the
internal energy of the plasma via multiple shocks. For comparison,
the combined X-ray luminosity of the NW ridge and SE clumps is  
$\sim 5 \times 10^{34}$ erg s$^{-1}$, whereas for the whole
bipolar structure the value will be perhaps a factor 2 higher 
(excluding the central few parsecs where the Sgr A East SNR dominates).
The observed X-ray emission thus represents only a tiny
fraction ($\sim 10^{-4}$) of the available wind energy. However,
the thermal energy stored within the NW ridge and SE clumps
amounts to $\sim 2 \times 10^{49} \rm~erg$ (assuming $f \approx 1$),
which is roughly half of their bulk kinetic energy if the outflow velocity
is $\sim 1000 \rm~km~s^{-1}$. For comparison the wind from the
Central Cluster releases a total energy of $\sim 
3 \times 10^{50}  \rm~erg$ over a timescale of 
of 13000 yr.  The implication is that the putative outflow
stores a substantial fraction of the energy deposited 
by the central wind into its surroundings.
 
An alternative explanation for the origin of the outflow
is that it is driven by intermittent outbursts on Sgr A*
(\citealt{markoff10}).  In this scenario,  Sgr A* remains in  a quiescent
state, apart from, on a rare occasions, when a star or cloud
passes close enough  to the SMBH to give rise to a
tidal disruption event.  The energy released
as matter is accreted onto the SMBH might then drive
a wind along the axis of a putative accretion disc, although
whether this would result in a bipolar configuration
aligned roughly perpendicular to the Galactic Plane
remains an open question. Alternatively the CND might again
provide the requisite collimation. To zeroth order, the X-ray
emitting clumps are spaced at radial
intervals of $\sim$1.7 arcmin (4 pc) -- see Fig. \ref{fig:cone}.
If we again assume an outflow velocity of 1000  km s$^{-1}$
then the implied time interval between successive tidal
disruption events is $\sim 4000$ yrs. This compares
with theoretical estimates of the tidal disruption rate of
$\sim 10^{-4}$ yr$^{-1}$ \citep{alexander05}. 

Previous X-ray studies of this region have not specifically
identified the arc-like feature which can be traced from the eastern
boundary of the Sgr A East SNR towards the south-east. The total extent
of this arc is at least 2.5 arcmin (6 pc). The morphology  and
orientation of the  arc suggests a possible connection 
with the outer SE clump. Spectral analysis (of region E) further
confirms that the X-ray emission is thermal in nature with similar
properties to the SE clumps. Unfortunately the origin of this arc
remains unclear.

\section{A new SNR candidate in the GC?}
\label{sec:NE}

\subsection{Images of the north-east enhancement}
\label{sec:imageNE}

Fig. \ref{fig:images}c shows the 2--4.5 keV image of the $6.3\arcminn
\times 6.3\arcmin$ (14.7 pc $\times$ 14.7 pc) region to the north-east
of Sgr A*, where the soft X-ray emission is notably enhanced. 
The brightest feature in the field coincides with an extended
X-ray source seen in \textit{Chandra} observations
which, according to \citet{wang02}, is a likely PWN. The underlying
neutron star may coincide with a \textit{Chandra} point source
designated as CXOGCS J174621.5-285256 (\citealt{wang02}), which
is located at the peak surface brightness 
in Fig. \ref{fig:images}c ({\it i.e.,} roughly 45 arcsec south of
an unrelated point X-ray source which is masked out in
this image).  The associated nebulosity of the PWN extends
up to 1.3 arcmin (3 pc) to the south-east (towards a
second unrelated masked source in the {\it XMM-Newton}
image).   This possible PWN, designated G0.13-0.11, has 
also been described both as an ``X-ray thread'' and as an 
``X-ray filament'' (\textit{e.g.}, \citealt{yusef02b}, 
\citeyear{yusef02a}; \citealt{johnson09}). 

Excluding the contribution of the PWN, the 2.0--4.5 keV emission in
this north-east region must be predominantly
thermal in nature as evidenced by the associated S$+$Ar line flux 
(Fig. \ref{fig:images}d).
The peak surface brightness in the S$+$Ar image coincides with the
south-eastern extension of the PWN, suggesting an association
between the PWN and the SN which presumably energised
the soft-thermal emission.  That said, the S$+$Ar surface brightness
remains high over the whole of the central and north-west segment
of the image, with a northern extent which can be traced as far
as the  Quintuplet Star Cluster (the masked source just visible
at the top edge of the image).

\begin{figure*}
\centering
\begin{tabular}{cc}
\includegraphics[width=65mm,angle=0]{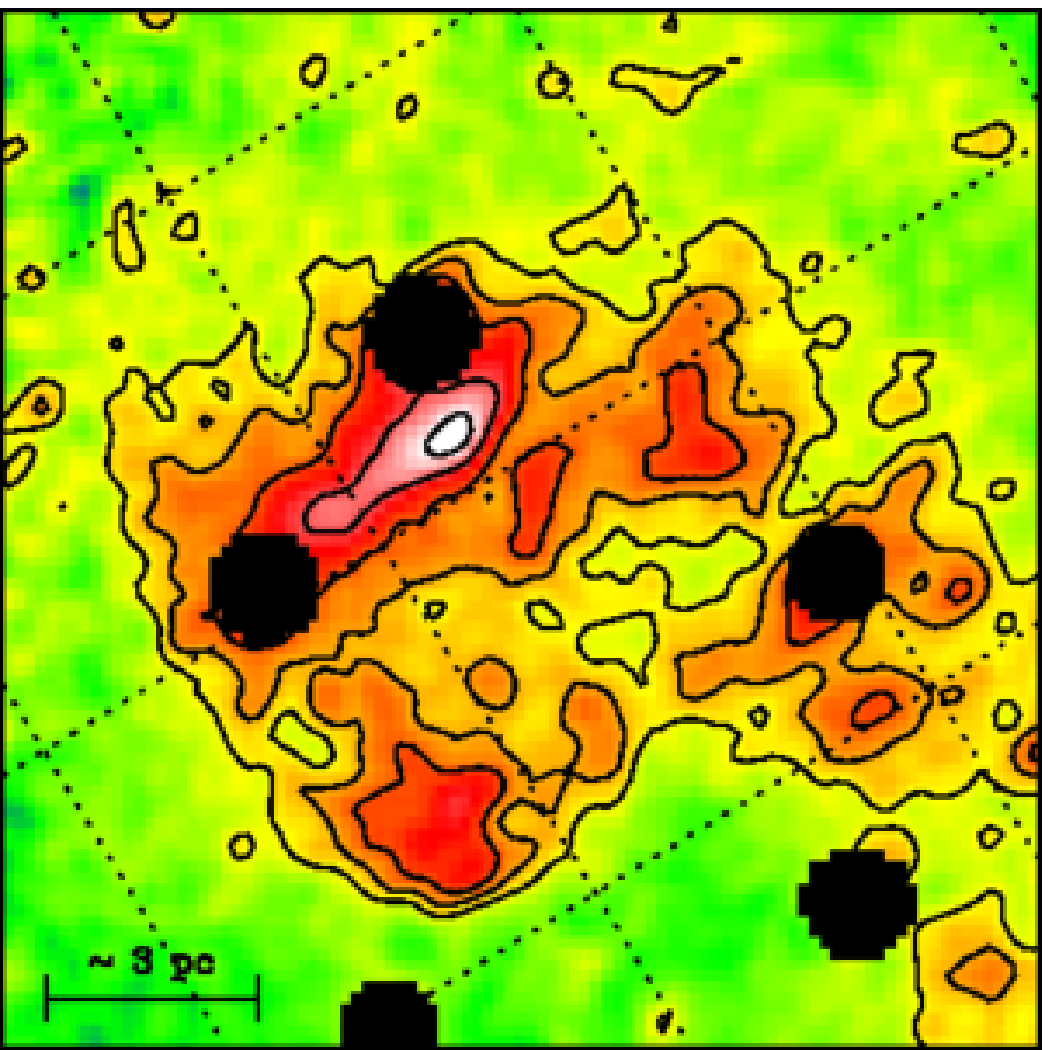}&
\includegraphics[width=65mm,angle=0]{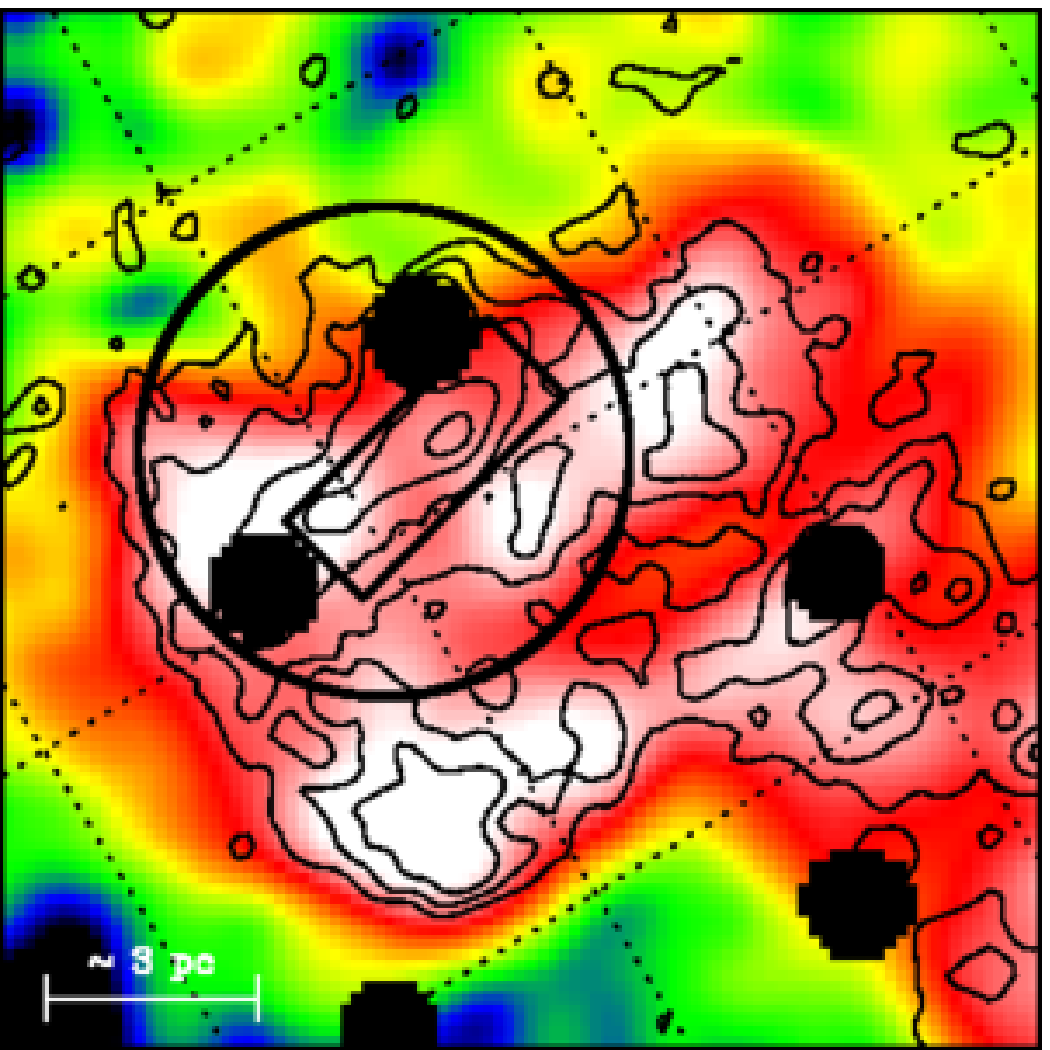}\\
\includegraphics[width=65mm,angle=0]{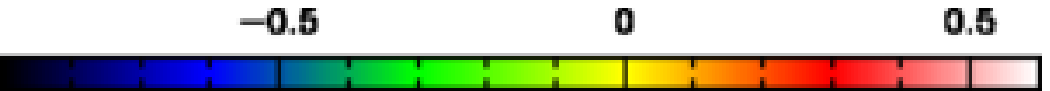}&
\includegraphics[width=65mm,angle=0]{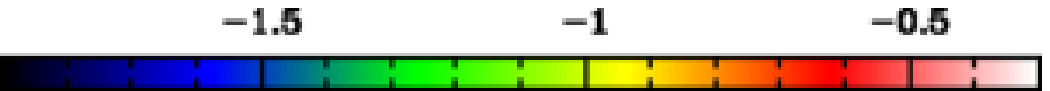}\\
\end{tabular}
\caption{Images    of    the   north-east  enhancement    covering
a $6.3\arcminn \times 6.3\arcmin$~ field.
In both cases, the intensity scaling is logarithmic as per the 
colour bar shown at the bottom  of the panel (where the units are
log$_{10}$ (count/20 ks/pixel)). 
The continuum images have been spatially smoothed using a Gaussian function
of width $\sigma$ = 4 arcsec, whereas for the Fe64 image $\sigma$ = 10 arcsec.
The black filled circles correspond to the regions excised by the
spatial mask. The dotted lines represent a Galactic coordinate grid with
2.4 arcmin spacing, \textit{Left-hand panel:} The broad-band 4.5--6 keV
image. \textit{Right-hand panel:} The 6.4-keV line image after subtracting
the contribution of the unresolved low-luminosity sources.
The black rectangle  represents the spectral extraction region
used for the  PWN  G0.13-0.11, whereas  the black  circle 
(excluding the rectangular region containing the PWN) defines the spectral
extraction region for the surrounding nebulosity.  Contours from the
4.5--6 keV image are superimposed.}
\label{fig:imageNE}
\end{figure*}

As noted by a number of authors, the PWN G0.13-0.11 lies in close proximity
(at least on the plane of the sky) to the molecular cloud, G0.11-0.11
(also known as the G0.13-0.13 cloud) (\citealt{tsuboi97}; \citealt{oka01};
\citealt{yusef02b}; \citealt{handa06}; \citealt{tsuboi11}). The position
of the PWN in relation to the dense molecular gas is illustrated in
Fig. \ref{fig:imageNE}, which shows both the
4.5--6 keV image and the Fe64 narrow-band image of this region.
The bright peak of the PWN (the likely location of the putative pulsar)
and the extension to the south-east are very prominent features in the
hard 4.5--6 keV band, but are not evident in the line image.
Setting aside the PWN emission, there is a strong correlation 
in Fig. \ref{fig:imageNE} between the 4.5--6 keV and Fe64
data, the former tracing the hard
continuum reflected from the dense clouds and latter tracing the
iron fluorescence which results from the same X-ray illumination
(\citealt{koyama07b}; \citeyear{koyama09}; \citealt{park04}; 
\citealt{ponti10}; \citealt{capelli12}). 
The northern edge of the G0.11-0.11 molecular cloud
overlaps with the PWN position, although it remains uncertain
whether the PWN is actually embedded in the cloud. An additional
link is with the set of narrow non-thermal filaments known as
the Radio Arc \citep{yusef84}, which are aligned at right-angles to
the Galactic Plane and extend over 15 arcmin
at  \textit{l}$_{\textsc{ii}} \approx 0.18\degn$. 
A curious aspect of the PWN morphology
is a (modest) curvature down its length which matches 
a bow-shaped radio feature located along the south-western
periphery of the Radio Arc region (\citealt{wang02}),
suggesting an interaction between the radio filament
and the PWN (\citealt{wang02}). It has further been suggested
that the collision of the G0.11-0.11 molecular cloud with
the enhanced magnetic field of the non-thermal filaments gives rise
to the acceleration of particles to relativistic energies;
these same particles, once injected, drift along the field lines
and hence go on to illuminate the full vertical structure of
the filaments (\citealt{tsuboi97}; \citealt{yusef02b};
\citealt{tsuboi11}).

In the present paper we focus on the possibility that the
region of highest surface brightness in the S$+$Ar image represents
the centre of an SNR.  Hereafter, we refer to this source as
SNR G0.13-0.12. In order to study the spectral properties
of this source, we have extracted spectra from
two regions - one encompassing the PWN and the other the
brightest region in the S$+$Ar image (but excluding the PWN
at its centre) -- see Fig. \ref{fig:images}d and Fig. \ref{fig:imageNE}.

\subsection{Spectral constraints}
\label{sec:specNE}

The presence of the PWN at the centre of the region in which the soft-thermal
emission reaches its peak surface brightness requires special accommodation
in the spectral fitting. Our approach was to extract spectra
separately from the PWN and SNR  regions identified above (noting that the
SNR region specifically excludes the PWN region which lies interior to it). 
The resulting spectra (hereafter referred to as the PWN and SNR spectra)
were fitted with a model comprising four emission components, namely the three
GC components described earlier (see \S\ref{sec:spectralx})
plus an absorbed power-law representing the continuum emission
of the PWN. For the latter the nominal free parameters were the
normalisation, photon-index $\Gamma$ and column density $N_{\rm{H}}$; however
in the event we found it necessary to tie the $N_{\rm{H}}$ applied to the
PWN emission to that pertaining to the soft-thermal emission component. 
In the spectral fitting, the PWN and SNR spectra were fitted jointly in
\textsc{xspec} with all the parameters, other than the emission component
normalisations, tied across the two datasets. 

The resulting best-fit value for the slope of the PWN continuum was
$\Gamma = 1.8^{+0.6}_{-0.5}$, fully consistent with the values
quoted by \citet{wang02} and \citet{johnson09}.
The total 2--10 keV unabsorbed flux
of  this power-law component was 1.4 $\times$ 10$^{-12}$ erg cm$^{-2}$ s$^{-1}$ (59 per cent of which was contained within the PWN extraction rectangle with the remainder spilling over into the SNR extraction circle). The corresponding
X-ray luminosity for a source at the GC distance is 1 $\times$ 10$^{34}$ erg s$^{-1}$, again broadly consistent with the published estimates.

The best-fitting parameters for the soft-thermal component
are listed in the fourth column of Table \ref{tab:bestfit}
(where the quoted normalisations and fluxes are the sum across
the SNR and PWN regions, thus representing a best estimate of
the total flux from the putative SNR). Fig. \ref{fig:spectra}
shows the corresponding best-fitting spectra and residuals for
the SNR region. The column density measured for the putative SNR (5.6$^{+0.5}_{-0.4} \times $10$^{22}$ cm$^{-2}$) is fully consistent with the fiducial GC $N_{\rm{H}}$ referred to earlier. Similarly the temperature of the plasma ($kT = 1.1^{+0.1}_{-0.1}$ keV) is very typical of the GC region (\S\ref{sec:cuts}).   The inferred abundances relative to solar of Si, S and Ar are all greater than 1, albeit with large errors; however, the weighted average over the 3 elements
is $1.75^{+0.21}_{-0.10}$, indicative of a modest over-abundance of these
elements in the hot plasma.

Using the best-fitting parameters from  Table  \ref{tab:bestfit}, we have
calculated the physical parameters of the X-ray emitting plasma
associated with the proposed SNR.
In this case the plasma volume was assumed to be spherical and of
radius 1.5 arcmin (3.5 pc).  The results are summarised in the
fourth column of  Table  \ref{tab:plasmax}.

The \texttt{vapec} thermal plasma model assumes that the plasma is
optically thin and in collisional ionization equilibrium (CIE). In
order to test the latter assumption, we replaced  the  \texttt{vapec}
component with  a  \texttt{vnei} component  and
re-fitted the  model\footnote{A Gaussian  line had  to be
included at 3.13  keV to account for the  Argon emission which appears 
not to be included in the \texttt{vnei} code.}. The ionization
parameter obtained was consistent with $\tau \sim$ 10$^{12}$ cm$^{-3}$ s (with a
lower  limit  of $\sim$ 7  $\times$  10$^{11}$  cm$^{-3}$ s). 
For an  electron density  of 1.2$f^{-1/2}$ cm$^{-3}$ (Table  
\ref{tab:plasmax}), 
this implies a lower limit on the ionization timescale of
1.8 $\times$ 10$^{4}$ yrs (assuming $f\sim1$).
In practice this is consistent with our original assumption of a CIE plasma. 

\subsection{Comments on the SNR hypothesis}
\label{sec:natureNE}

The peak in  the X-ray emission for SNR  G0.13-0.12 coincides with
the southern extension of PWN G0.13-0.11. However, the non-thermal
emission from  this PWN  accounts for  only $\sim 44$  per cent  of the
total emission. SNR G0.13-0.12  appears to be a centrally-concentrated
source of  thermal X-ray emission consistent  with a thermal-composite
description\footnote{This   class   of   SNR   are   also   known   as
mixed-morphology     remnants     (\citealt{rho98};     see     also
\citealt{vink12}).}. In thermal-composite SNRs the bright thermal X-ray
core is typically accompanied by a shell-like radio morphology,
whereas in the present case, there is little concrete evidence for
a radio shell or even a partial shell. However, SNR G0.13-0.12 lies
within a complex region containing several radio-bright features
such as the Radio Arc \citep{yusef84},  the Arched filaments
\citep{morris89} and the Sickle \citep{lang97}, plus a myriad of
fainter structures ({\it e.g.}, \citealt{reich03}). Hence,
the presence of an associated radio shell may, in this instance,
be masked. SNR G0.13-0.12 is, in fact, spatially coincident  with a
structure known as the ``Bubble'' which  is a low-contrast feature
in 21 cm  radio continuum images \citep{levine99},  but appears  as
a prominent cavity in the distribution of warm dust as revealed by
21 $\mu$m continuum data 
(\citealt{egan98}; \citealt{price01}). \citet{simpson07} 
suggest that the source of the dust excitation is likely to be
the Quintuplet Cluster, although the cluster is well
offset from the centre  of the Bubble. In the  present
context, it is possible to conjecture either that the dust cavity was 
formed through  the interaction  of SNR G0.13-0.12 with the surrounding
ISM or that SNR G0.13-0.12 exploded into an pre-existing cavity.
As noted earlier, in the region of SNR G0.13-0.12, the bright
soft X-ray emission extends well beyond the central core and, in
particular, can be traced in the north-west sector as far as 
the Quintuplet Cluster; it is plausible that this is the result of the
breakout of the soft thermal plasma from the SNR into a pre-existing
cavity.

From Table \ref{tab:plasmax}, the internal thermal energy of SNR G0.13-0.12
is $\sim3  \times 10^{49}$ erg  (assuming $f \sim1$)  corresponding to
around  3 per cent  of  the  total explosive  energy of a SN.
Furthermore,  the  mass  contained  within  the
remnant is $\sim5\Msunn$, which is  not excessive in the context
of the expected ejecta mass from  a type II SN explosion
\citep{vink12}.  Therefore, the interpretation of the enhanced
soft thermal emission in this region in terms of a single SN
event is credible. 
The derived  plasma temperature ($kT \approx 1.1$ keV) is 
similar to those observed from other thermal-composite SNRs,   
\textit{e.g.}, IC 443 ($kT \approx 1$ keV,
\citealt{kawasaki02}), W44  ($kT  \approx 0.84$ keV,
\citealt{kawasaki05}) and Kes 79 ($kT \approx 1.3$ keV,
\citealt{rho98}).
Earlier we noted from the spectral analysis
that the  assumption of CIE is valid with an inferred plasma
ionization timescale  of at least $1.8 \times 10^{4}$ yr. 
This is consistent with a thermal-composite classification,
since an age of 20 000 yr would seem to be an  ubiquitous
characteristic of this type of  remnant (\citealt{vink12}).
Many thermal-composite SNRs show evidence for metal-rich thermal
plasmas (see for example, \citealt{lazendic06}), which is also
in line with our measurement of an overabundance of
several alpha elements in the X-ray spectrum of G0.13-0.12. 

In thermal-composite remnants the bright thermal X-ray emission may 
arise when dense cloudlets, which  survive the passage of the forward
shock from the SN explosion, slowly evaporate and enhance the
density of the hot interior  (\citealt{white91}).  In an  alternative model,
\citet{cox99} suggest that as the forward shock decelerates to velocities below
$\sim  200  \rm~km~s^{-1}$, strong  cooling sets in  which diminishes
the  X-ray emission  from the SNR  shell. However, the hot thermal 
X-ray emission from  the interior is maintained as the  result of
thermal conduction \citep{cui92} and turbulent mixing processes 
(\citealt{shelton99}).

As  mentioned  previously, SNR G0.13-0.12 is positioned on the northern
boundary of the molecular cloud G0.11-0.11. It has been suggested that
an association  with  molecular   clouds   may   represent  a   defining
characteristic  of thermal-composite remnants \citep{rho98}. 
The presence of dense clouds is in fact a critical ingredient of
the cloud-evaporation model (\citealt{white91}), that has been
successfully applied to thermal-composite remnants such as
W28 and 3C  400.2 \citep{long91} and  W44 \citep{rho94}.  
If we assume (na\"{i}vely) that SNR G0.13-0.12 is in the  late Sedov phase,
we  can determine the density of the medium into which it is expanding
as,  $n  \sim (14/R)^{5}t^{2}E_{0}$, where $n$ is the density in cm$^{-3}$,
$R$ is the remnant radius in pc, $t$ is its age in units of $10^{4}$ yr
and $E_{0}$ is the explosion energy in units of  $10^{51}$ erg.
Setting $R=7$ (roughly twice the radius of the centrally bright
X-ray emission - see figure 1 in \citealt{rho98}), $t=2$ and $E_{0}=1$,
we obtain $n \sim 120 $  cm$^{-3}$ suggesting that the  progenitor  of
SNR G0.13-0.12 may have been embedded within a diffuse cloud
environment but not within in a dense molecular cloud
with  $n \sim 10^{4} $ cm$^{-3}$.

The SNR G0.13-0.12 and the molecular cloud G0.11-0.11 lie within the error
box of the EGRET $\gamma$-ray source 3EG J1746-2851 (\citealt{hartman99},
see also \citealt{yusef03a}).  The greatly improved sensitivity and spatial
resolution afforded by the  \textit{Fermi}  LAT, has lead to
the detection of multiple GeV sources in the GC  (\citealt{abdo10c}; 
\citealt{nolan12}) with the source source 2FGL J1746.4-2851c positionally
coincident  with SNR G0.13-0.12 (see also \citealt{yusef13}). A number of
other $\gamma$-ray sources have been linked with thermal-composite
SNRs residing near molecular clouds,  including W51C \citep{abdo09},
W44 \citep{abdo10a} and IC 443 \citep{abdo10b}. The  current
understanding is that protons and nuclei, accelerated in
the SNR shock fronts, subsequently diffuse out of the  nebula
and  interact  with   nearby  molecular  material to produce
neutral pions  \citep{aharonian94}, the decay of which gives 
rise to the $\gamma$-ray emission.  An VHE counterpart has yet 
to be unequivocally detected with H.E.S.S. However, the presence
of the nearby source, HESS J1745-290, which is thought to be
associated with either Sgr A* or PWN G359.95-0.04 (\citealt{acero10},
and references therein) gives rise to significant source confusion
in the region of the SNR. However, when the contribution of
HESS J1745-290 is subtracted, bright residual
TeV emission is evident overlapping the location of SNR G0.13-0.12
and the likely associated molecular cloud \citep{aharonian06}.

\section{A Superbubble in the GC?}
\label{sec:sb}

\subsection{Images of the region to the south of Sgr A*}
\label{sec:imagesb}

Fig. \ref{fig:images}e shows the 2--4.5 keV broad-band image of a
region to the immediate south of Sgr A* of size $23\arcminn \times
21\arcmin$ (54 pc $\times$ 49 pc).
Over much of this field the average surface brightness
is less than a third of that pertaining in the north-east region considered 
earlier. In this southern region, a loop-like structure is apparent which may
delineate a potential Galactic superbubble (\citealt{mori08}, 
\citeyear{mori09}). On the basis of \textit{Suzaku} observations, 
\citet{mori09} define the outer and inner bounds of the loop feature
by two ellipses, both centred on ($l_{\textsc{ii}}$, $b_{\textsc{ii}}$) = 
(+359.8312\degn, -0.1367\degn), with dimensions 
$20\arcminn \times 16\arcmin$~ 
(46 pc $\times$ 37 pc)  and $8\arcminn \times 6.4\arcmin$ 
(18.5 pc $\times$ 15 pc), respectively.  As is evident from
Fig. \ref{fig:images}e, the northern part of this
structure overlaps the Sgr A East SNR and the south-east lobe of the
bipolar outflow, which together provide a very bright backdrop against
which it is not possible to discern putative superbubble emission. 
Outside of
this confused region, the loop structure is reasonably complete, although
the south-eastern and western segments are particularly well traced by
relatively bright emission. 
The brightest regions apparent in the 2--4.5 keV image
correspond to local peaks in the S$+$Ar line image 
(Fig. \ref{fig:images}f), indicative of the fact that these represent
concentrations of hot thermal plasma. Below we investigate the X-ray
spectra extracted from the two regions defined by the (small) ellipses
in the line image and designated, respectively, as
G359.77-0.09 and G359.79-0.26 (see \citealt{senda03}).

\subsection{Spectral constraints}
\label{sec:specsb}

As noted earlier, the G359.79-0.26 region lies outside of the field of
view of our standard spectral dataset (0202670801) and hence for
this source we have been forced to use an alternative observation
(0112971001), with a much reduced exposure time. 

We utilise the same spectral model as employed previously to study the
bipolar outflow  (\S\ref{sec:specbi}), namely  one which  accounts for
both  the   underlying  emission   from  unresolved-sources   and  the
soft-thermal emission  in the  region. As  usual, we  assume initially
that the latter can be  adequately represented by a single-temperature
\texttt{vapec} component in \textsc{xspec}.

The  two-component model  provides a  good fit to the spectra of
both sources. As expected
the normalisation of the bremsstralung component representing the
unresolved sources is higher for G359.77-0.09 than G359.79-0.26, 
commensurate with the location of the former closer to the Galactic Plane. 
The resulting best-fitting parameters pertaining to the soft-thermal
plasma emission are reported in the last two
columns of Table  \ref{tab:bestfit}. Fig. \ref{fig:spectra}
shows the corresponding best-fitting spectra and residuals.

The derived temperature for G359.077-0.09 is somewhat lower
than that of G359.79-0.26 (approximately 0.7 keV versus 1.0 keV),
whereas the $N_{\rm{H}}$ values show the opposite trend (5.9 $\times$ 10$^{22}$ cm$^{-2}$ versus 4.4 $\times$ 10$^{22}$ cm$^{-2}$). These measurements are in excellent agreement with published results (\citealt{mori08}, \citeyear{mori09}). The column-density measurements place the loop structure, unequivocally,
within the GC region (\citealt{mori09}). The derived abundances for Si, S and Ar are, in the main, close to solar values and again very consistent with
the published results. Interestingly, the hint of a slight over-abundance
of S and Ar relative to solar in the {\it XMM-Newton} data for 
G359.79-0.26 is also present in the \textit{Suzaku}
measurements.

We   have  again   used   the  best-fitting   parameters  from   Table
\ref{tab:bestfit} to  determine the  physical parameters of  the X-ray
emitting plasma within G359.77-0.09  and G359.79-0.26.  We assumed the
plasma to be contained within ellipsoids of dimension 11.1 pc $\times$
5.0 pc  $\times$ 5.0 pc  and 11.1 pc $\times$  5.5 pc $\times$  5.5 pc,
respectively,  and  then  applied  the same  methodology  as  discussed
earlier.   The  results  are  given  in  last  two  columns  of  Table
\ref{tab:plasmax}.

Finally, we have tried the  experiment of replacing the \texttt{vapec}
component with the  \texttt{vnei} model in order to  test the validity
of the assumption  that the soft X-ray emitting plasma  is in CIE. For
both  G359.77-0.09  and  G359.79-0.26   we  found  that  the  inferred
ionization parameter, $\tau$  was of the order  of $10^{12}$ cm$^{-3}$
s, consistent with our original assumption of CIE.
For an  electron density in the range $ (0.7-1.1)$ 
cm$^{-3}$ (Table  \ref{tab:plasmax}), 
this implies a lower limit on the ionization timescale ranging
from 2.9--4.5 $\times$ 10$^{4}$ yr (assuming $f\sim1$).

\subsection{Comments on the superbubble hypothesis}
\label{sec:natureSB}

Superbubbles are cavities in  the ISM filled with hot plasma,
which are formed as a result of  sequential  co-located SN explosions 
(\citealt{berkhuijsen71}; \citealt{elmegreen77}) and/or the combined
effect of the stellar winds of an association of OB stars  
(\citealt{castor75}; \citealt{weaver77}). 

\citet{mori09} have  recently proposed that the enhanced soft X-ray 
surface brightness within the regions designated as G359.77-0.09
and G359.79-0.26 represent substructure within the extended shell
of a superbubble located near to the GC. The argument in
favour of this hypothesis revolves around both the morphology
of the structure and the fact that the total thermal energy contained
within the hot plasma shell is $ \sim 10^{51} f^{1/2}$ erg.
Assuming a filling factor $f$ close to unity and that 
a maximum of 10 per cent of the $10^{51}$ erg liberated in a 
SN explosion can be converted to thermal energy \citep{vink12},
this implies the need for multiple SN events.

The present results from {\it XMM-Newton} confirm that a
near-complete X-ray shell can be traced southward of Sgr A*
over an extent of  roughly 45 pc. The {\it XMM-Newton}
observations also confirm the thermal nature of this structure
and the general properties of the bright sub-regions. The
thermal energy contained within the G359.77-0.09 and
G359.79-0.26 features are further estimated to be  $1.2 \times 10^{50} f^{1/2}$ erg
and $1.3 \times 10^{50} f^{1/2}$ erg, respectively. On the basis of these
measurements, one can conjecture that these two features might be the result
of the passage of the blast wave from separate SN explosions through
highly clumped regions of interstellar
material (resulting in $f < 1$), and that the shell structure is not a coherent
physical entity. However, the apparent limb-brightened characteristics
of these two features in relation to the putative loop structure
is a persuasive counter-argument.

From  the spectral fitting, the ionization timescales, $t_{ion}$, 
for G359.77-0.09 and  G359.79-0.26 were estimated to be at least 30 000 yr
(assuming $f \approx 1$).  For comparison the estimated cooling timescales
listed in Table \ref{tab:plasmax} are two orders of magnitude longer and
imply that radiative cooling is not a significant factor.  To a rough
approximation we may assume that the superbubble structure is in a late
Sedov  phase of evolution. Taking the  radius of
the superbubble to be $\sim20$ pc (the average of the semi-major
and semi-minor axes of the outer edge of the shell) and the age
to be 30 000 yr, then the Sedov solution gives the density of the medium
into which the most recent supernova blast wave has expanded to
be $\sim 1 $ cm$^{-3}$. In the GC context this suggests the presence
of a pre-formed cavity, perhaps excavated by earlier SN. However, the same Sedov
solution predicts a plasma temperature $\lesssim 0.1$ keV, which
is at variance with the measured values of 0.7 keV and 1.1 keV
for G359.77-0.09 and  G359.79-0.26, respectively. Clearly, in these circumstances a more sophisticated model describing the evolution of a supernova explosion within a pre-formed bubble is required (\citealt{vink12}; and references therein). In the present context, it would seem likely that the X-ray bright components represent relatively dense regions within the confines of the superbubble loop structure in which recent shock heating has given rise to an enhanced X-ray luminosity ({\it e.g.}, \citealt{dwarkadas05}).

The central 100-pc region hosts an array of dense clouds with the potential to leave absorption imprints in soft X-ray images and spectra. The best-studied molecular complex within the bounds of the putative superbubble is the so-called 20 km s$^{-1}$ cloud, which is a very prominent feature in molecular-line maps ({\it e.g.}, \citealt{oka98}; \citealt{tsuboi99}; \citeyear{tsuboi11}). This cloud is also clearly seen in emission in the {\it Herschel} 250 \micron~image of the GC and, by virtue of its high cold-dust opacity, appears silhouetted against the warmer 70 \micron~background in the dust temperature map derived from the {\it Herschel} data (\citealt{molinari11}). The 20 km s$^{-1}$ cloud (also known as GCM -0.13-0.08) extends over much of the northern half  of the superbubble interior and with a peak molecular hydrogen column density of approximately $6.7 \times 10^{23} \rm cm^{-2}$ (\citealt{tsuboi11}) would certainly shadow any background soft thermal X-ray emission. In principle, therefore, the apparent deficit of soft X-ray emission from the superbubble interior might be a consequence of a particular pattern in foreground absorption. However, the observed X-ray emission, in particular the morphology of the G359.77-0.09 feature, shows little evidence that it is directly shaped by absorption in the 20 km s$^{-1}$ cloud (comparing, for example, the SiO $v = 0, J=2-1$ images in figure 9 of \citet{tsuboi11} with the current X-ray maps). A ridge-like molecular feature seen in CS $J=1-0$ images (\citealt{tsuboi99}), in the velocity range -30 -- 0 km s$^{-1}$,  similarly runs roughly north-south along the inner edge of the G359.77-0.09 X-ray emission. This molecular feature further extends beyond the southern bound of the superbubble towards the Sgr C region.  Soft X-ray absorption arising in this molecular component might explain the gap in the superbubble loop at its southern boundary, but again the evidence for this conjecture is at best tentative.

As previously noted by  \citet{mori09}, the known superbubbles
within our own galaxy, typically have very soft X-ray spectra
and much larger spatial scales than this GC feature.
For  example,   the Gemini-Monoceros superbubble has a dimension of
roughly 140 pc and an X-ray plasma temperature  $kT\approx0.2$  keV 
\citep{plucinsky96} and the Cygnus superbubble has an extent of
roughly 450 pc also with an X-ray plasma temperature $kT\approx0.2$ 
keV \citep{guo95}. However, recent observations of the LMC have led to the
discovery of superbubbles with temperatures $kT \sim 1$ keV 
({\it e.g.,} \citealt{sasaki11}; \citealt{kavanagh12}). 

A further problem with the GC superbubble interpretation is 
that there is no known OB association  that
can be identified as the likely origin of the SN progenitors;
however, given the high obscuration towards the GC, 
current catalogues  may be far from complete.
There is also a lack of convincing evidence for the loop 
structure in current radio observations, albeit 
to the north, the Sgr A radio source provides a bright
backdrop against which it would be difficult to trace
faint radio emission\footnote{There is a feature in
the 90 cm radio data of \citet{larosa00} which
extends from the southern boundary of the Sgr A radio
complex down towards G359.79-0.26, but any
association with the putative superbubble loop
looks at best tenuous.}. 

Unfortunately, at present, the evidence supporting the interpretation of
the apparent X-ray loop to the south of Sgr A region in terms of a
new GC superbubble remains rather inconclusive.

\section{Summary and Conclusions}
\label{conclusions}

In this paper, we have used \textit{XMM-Newton}  observations to create
a mosaiced X-ray image of the central 100-pc region of the GC in a
variety of bands encompassing both the broad 1--10 keV range and several narrow
bands centred on the prominent K-shell lines such as those of helium-like
silicon, sulphur, argon and iron.
We use these images in combination with X-ray spectral data to  investigate
the  nature of  the diffuse soft thermal X-ray emission pervading the region. 
From latitudinal cuts through the various images, we find that the bulk of
the soft X-ray emission is thermal in nature with a plasma temperature of 
$kT \approx 1$  keV, subject to absorption  by a  column density of 
$N_{\rm{H}} \approx  6 \times  10^{22}$ cm$^{-2}$. Deviations from this norm,
are seen in  some places, most notably in the region offset between 
$2\arcmin-10\arcmin$ to the Galactic south of Sgr A*, where the plasma temperature is enhanced,
resulting in the presence of additional helium-like iron line emission
over and  above that attributable to the unresolved  source population
studied in Paper I.

Three distinct large-scale features are seen in the 2--4.5 keV
image and we investigate the properties of these in some detail.
The first is an extended band of emission centred  on  Sgr  A*
and aligned  perpendicular  to  the  Galactic  Plane, which is
theorised  to be a bipolar outflow. We  discuss the  origin of  this
bipolar morphology and suggest that the plasma is produced through the
shock heating of  high-velocity winds emanating from  massive stars in
the Central Cluster.  The resulting outflow may then be collimated by the
Circumnuclear   Disc.  One   problem   with  this   scenario  is   the
sub-structure  observed  in  the   feature,  which  cannot  be  easily
explained if the flow is quasi-continuous.  Alternatively,  the
outflow may  be driven by outbursts on Sgr A* as a result
of  tidal-disruption  events occurring at a rate of
roughly one event per 4000 yr. 

The second feature is the region of enhanced X-ray brightness situated
to the north-east of Sgr A*, which is centred on the location of 
the candidate PWN G0.13-0.11.  We suggest that the centrally-concentrated
thermal X-ray emission represents the core of a thermal-composite
SNR. This  X-ray source lies (on the plane of the sky) at the edge of the
dense molecular cloud G0.11-0.11 and it seems likely that the SNR and
molecular cloud have some association.  The coincidence of
a \textit{Fermi} point source is perhaps indicative of cosmic-ray
hadrons accelerated in the SNR, interacting with the dense molecular
material and via $\pi^{0}$ decay giving rise to the $\gamma$-ray
emission.  This combination of characteristics are commonly
observed in thermal-composite SNRs.

Finally, we focus  on the extended loop feature to the  south of
Sgr A*,   which   has  previously  been   interpreted
as the shell of a GC superbubble.   Our \textit{XMM-Newton}  observations
are not inconsistent with this scenario, but unfortunately the evidence
in support of this being a coherent physical structure near to the
GC remains somewhat tenuous.

\section*{Acknowledgments}

This work is based on \textit{XMM-Newton} observations, an ESA mission
with  instruments  and contributions  directly  funded  by ESA  member
states  and the  USA  (NASA). VH  acknowledges  the financial  support
provided by the UK STFC research council.

\bibliography{soft_thermal_refs}
\bibliographystyle{mn2e}

\bsp

\label{lastpage}

\end{document}